\batchmode
\makeatletter
\def\input@path{{"/home/jacob/Documents/Work/My Papers/2023-Stochastic Processes and Quantum Theory/"}}
\makeatother
\documentclass[11pt,twoside,english]{article}
\usepackage[T1]{fontenc}
\usepackage[latin9]{inputenc}
\usepackage{color}
\definecolor{lyxboxbgcolor}{rgb}{0.980469, 0.941406, 0.902344}
\usepackage{babel}
\usepackage{mathtools}
\usepackage{amsmath}
\usepackage{amssymb}
\usepackage{geometry}
\usepackage{setspace}
\usepackage[pdftex,pdfusetitle,
 bookmarks=false,
 breaklinks=false,pdfborder={0 0 0},pdfborderstyle={},backref=false,colorlinks=false]
 {hyperref}

\makeatletter

\usepackage{endnotes}

\def\smalloverbrace#1{\mathop{\vbox{\m@th\ialign{##\crcr%
      \noalign{\kern3\p@}%
      \tiny\downbracefill\crcr\noalign{\kern3\p@\nointerlineskip}%
      $\hfil\displaystyle{#1}\hfil$\crcr}}}\limits}

\def\smallunderbrace#1{\mathop{\vtop{\m@th\ialign{##\crcr
   $\hfil\displaystyle{#1}\hfil$\crcr
   \noalign{\kern3\p@\nointerlineskip}%
   \tiny\upbracefill\crcr\noalign{\kern3\p@}}}}\limits}

\DeclareMathAlphabet{\mymathbb}{U}{bbold}{m}{n}

\makeatother

\begin{document}
\title{Quantum Systems as Indivisible Stochastic Processes}
\author{Jacob A. Barandes\thanks{Departments of Philosophy and Physics, Harvard University, Cambridge, MA 02138; jacob\_barandes@harvard.edu; ORCID: 0000-0002-3740-4418}
\thanks{This is the accepted manuscript, published in \emph{Studies in History and Philosophy of Science} (DOI: 10.1016/j.shpsa.2026.102213)}
\thanks{Some of this material in this paper was originally part of arXiv:2302.10778 (Version 1).}}
\date{}

\maketitle

\begin{abstract}
According to the stochastic\textendash quantum correspondence, a quantum
system can be understood as a stochastic process unfolding in an old-fashioned
configuration space based on ordinary notions of probability and \textquoteleft indivisible\textquoteright{}
stochastic laws, which are a non-Markovian generalization of the laws
that describe a textbook stochastic process. The Hilbert spaces of
quantum theory and their ingredients, including wave functions, can
then be relegated to secondary roles as convenient mathematical appurtenances.
In addition to providing an arguably more transparent way to understand
and modify quantum theory, this indivisible-stochastic formulation
may lead to new possible applications of the theory. This paper initiates
a deeper investigation into the conceptual foundations and structure
of the stochastic\textendash quantum correspondence, with a particular
focus on novel forms of gauge invariance, dynamical symmetries, and
Hilbert-space dilations.

\end{abstract}
\begin{center}
\global\long\def\quote#1{``#1"}%
\global\long\def\apostrophe{\textrm{'}}%
\global\long\def\slot{\phantom{x}}%
\global\long\def\eval#1{\left.#1\right\vert }%
\global\long\def\keyeq#1{\boxed{#1}}%
\global\long\def\importanteq#1{\boxed{\boxed{#1}}}%
\global\long\def\given{\vert}%
\global\long\def\mapping#1#2#3{#1:#2\to#3}%
\global\long\def\composition{\circ}%
\global\long\def\set#1{\left\{  #1\right\}  }%
\global\long\def\setindexed#1#2{\left\{  #1\right\}  _{#2}}%

\global\long\def\setbuild#1#2{\left\{  \left.\!#1\,\right|\,#2\right\}  }%
\global\long\def\complem{\mathrm{c}}%

\global\long\def\union{\cup}%
\global\long\def\intersection{\cap}%
\global\long\def\cartesianprod{\times}%
\global\long\def\disjointunion{\sqcup}%

\global\long\def\isomorphic{\cong}%

\global\long\def\setsize#1{\left|#1\right|}%
\global\long\def\defeq{\equiv}%
\global\long\def\conj{\ast}%
\global\long\def\overconj#1{\overline{#1}}%
\global\long\def\re{\mathrm{Re\,}}%
\global\long\def\im{\mathrm{Im\,}}%

\global\long\def\transp{\mathrm{T}}%
\global\long\def\tr{\mathrm{tr}}%
\global\long\def\adj{\dagger}%
\global\long\def\diag#1{\mathrm{diag}{\left(#1\right)}}%
\global\long\def\dotprod{\cdot}%
\global\long\def\crossprod{\times}%
\global\long\def\Probability#1{\mathrm{Prob}{\left(#1\right)}}%
\global\long\def\Amplitude#1{\mathrm{Amp}{\left(#1\right)}}%
\global\long\def\cov{\mathrm{cov}}%
\global\long\def\corr{\mathrm{corr}}%

\global\long\def\absval#1{{\left\vert #1\right\vert }}%
\global\long\def\expectval#1{{\left\langle #1\right\rangle }}%
\global\long\def\op#1{\hat{#1}}%

\global\long\def\bra#1{{\left\langle #1\right|}}%
\global\long\def\ket#1{{\left|#1\right\rangle }}%
\global\long\def\braket#1#2{{\left\langle {\left.\!#1\right|}#2\right\rangle }}%

\global\long\def\parens#1{(#1)}%
\global\long\def\bigparens#1{\big(#1\big)}%
\global\long\def\Bigparens#1{\Big(#1\Big)}%
\global\long\def\biggparens#1{\bigg(#1\bigg)}%
\global\long\def\Biggparens#1{\Bigg(#1\Bigg)}%
\global\long\def\bracks#1{[#1]}%
\global\long\def\bigbracks#1{\big[#1\big]}%
\global\long\def\Bigbracks#1{\Big[#1\Big]}%
\global\long\def\biggbracks#1{\bigg[#1\bigg]}%
\global\long\def\Biggbracks#1{\Bigg[#1\Bigg]}%
\global\long\def\curlies#1{\{#1\}}%
\global\long\def\bigcurlies#1{\big\{#1\big\}}%
\global\long\def\Bigcurlies#1{\Big\{#1\Big\}}%
\global\long\def\biggcurlies#1{\bigg\{#1\bigg\}}%
\global\long\def\Biggcurlies#1{\Bigg\{#1\Bigg\}}%
\global\long\def\verts#1{\vert#1\vert}%
\global\long\def\bigverts#1{\big\vert#1\big\vert}%
\global\long\def\Bigverts#1{\Big\vert#1\Big\vert}%
\global\long\def\biggverts#1{\bigg\vert#1\bigg\vert}%
\global\long\def\Biggverts#1{\Bigg\vert#1\Bigg\vert}%
\global\long\def\Verts#1{\Vert#1\Vert}%
\global\long\def\bigVerts#1{\big\Vert#1\big\Vert}%
\global\long\def\BigVerts#1{\Big\Vert#1\Big\Vert}%
\global\long\def\biggVerts#1{\bigg\Vert#1\bigg\Vert}%
\global\long\def\BiggVerts#1{\Bigg\Vert#1\Bigg\Vert}%
\global\long\def\ket#1{\vert#1\rangle}%
\global\long\def\bigket#1{\big\vert#1\big\rangle}%
\global\long\def\Bigket#1{\Big\vert#1\Big\rangle}%
\global\long\def\biggket#1{\bigg\vert#1\bigg\rangle}%
\global\long\def\Biggket#1{\Bigg\vert#1\Bigg\rangle}%
\global\long\def\bra#1{\langle#1\vert}%
\global\long\def\bigbra#1{\big\langle#1\big\vert}%
\global\long\def\Bigbra#1{\Big\langle#1\Big\vert}%
\global\long\def\biggbra#1{\bigg\langle#1\bigg\vert}%
\global\long\def\Biggbra#1{\Bigg\langle#1\Bigg\vert}%
\global\long\def\braket#1#2{\langle#1\vert#2\rangle}%
\global\long\def\bigbraket#1#2{\big\langle#1\big\vert#2\big\rangle}%
\global\long\def\Bigbraket#1#2{\Big\langle#1\Big\vert#2\Big\rangle}%
\global\long\def\biggbraket#1#2{\bigg\langle#1\bigg\vert#2\bigg\rangle}%
\global\long\def\Biggbraket#1#2{\Bigg\langle#1\Bigg\vert#2\Bigg\rangle}%
\global\long\def\angs#1{\langle#1\rangle}%
\global\long\def\bigangs#1{\big\langle#1\big\rangle}%
\global\long\def\Bigangs#1{\Big\langle#1\Big\rangle}%
\global\long\def\biggangs#1{\bigg\langle#1\bigg\rangle}%
\global\long\def\Biggangs#1{\Bigg\langle#1\Bigg\rangle}%

\global\long\def\vec#1{\mathbf{#1}}%
\global\long\def\vecgreek#1{\boldsymbol{#1}}%
\global\long\def\idmatrix{\mymathbb{1}}%
\global\long\def\projector{P}%
\global\long\def\permutationmatrix{\Sigma}%
\global\long\def\densitymatrix{\rho}%
\global\long\def\krausmatrix{K}%
\global\long\def\stochasticmatrix{\Gamma}%
\global\long\def\lindbladmatrix{L}%
\global\long\def\dynop{\Theta}%
\global\long\def\timeevop{U}%
\global\long\def\hadamardprod{\odot}%
\global\long\def\tensorprod{\otimes}%

\global\long\def\inprod#1#2{{\left\langle #1,#2\right\rangle }}%
\global\long\def\normket#1{\left\Vert #1\right\Vert }%
\global\long\def\hilbspace{\mathcal{H}}%
\global\long\def\samplespace{\Omega}%
\global\long\def\configspace{\mathcal{C}}%
\global\long\def\phasespace{\mathcal{P}}%
\global\long\def\spectrum{\sigma}%
\global\long\def\restrict#1#2{\left.#1\right\vert _{#2}}%
\global\long\def\from{\leftarrow}%
\global\long\def\statemap{\omega}%
\global\long\def\degangle#1{#1^{\circ}}%
\global\long\def\trivialvector{\tilde{v}}%
\global\long\def\eqsbrace#1{\left.#1\qquad\right\}  }%
\par\end{center}

\section{Introduction\label{sec:Introduction}}

\textquoteleft Indivisible\textquoteright{} stochastic processes are
a very new idea. They first appeared in the research literature in
a 2021 review article, which introduced them in passing as a generalization
of the sorts of stochastic processes described in textbooks (Milz,
Modi 2021, Fig. 6)\nocite{MilzModi:2021qspaqnp}. One can find rudimentary
notions of indivisible dynamics in earlier work on open quantum systems,
although those forms of indivisibility corresponded to quantum channels,
not to stochastic processes (Wolf, Cirac 2008)\nocite{WolfCirac:2008dqc}.\footnote{Divisibility and indivisibility in the sense of a stochastic process
should also not be confused with the older and altogether different
notion of \emph{infinite divisibility} for a static probability distribution.}

Like a textbook stochastic process, an indivisible stochastic process
is based on ordinary probability theory, and describes a system's
trajectory unfolding in an old-fashioned configuration space according
to probabilistic dynamical laws. However, as the present work will
review, indivisible stochastic processes do not satisfy the Markov
property, nor are they non-Markovian only in the familiar sense of
requiring the specification of higher-order conditional probabilities.
As a preview of the detailed discussion ahead, there are two key
differences between an indivisible stochastic process and a textbook,
non-Markovian stochastic process with \textquoteleft memory\textquoteright :
defining an indivisible stochastic process requires specifying a much
smaller and simpler set of underlying dynamical laws, and those underlying
dynamical laws are not generically decomposable or \textquoteleft divisible\textquoteright{}
over time (Barandes 2025)\nocite{Barandes:2025tsqc}.

Remarkably, indivisible stochastic processes generically exhibit all
the hallmark empirical features of quantum systems, including interference,
decoherence, entanglement, and noncommutative observables. These features
agree with the observed behavior of quantum systems not only qualitatively,
but quantitatively. Indeed, by a straightforward choice of dynamics,
an indivisible stochastic process can replicate all the empirical
predictions made using a given quantum system, while evading the famous
\emph{measurement problem} and the more recently named \emph{category problem}
(Ibid.)\nocite{Barandes:2025tsqc}. The measurement problem refers
to the manifest ambiguity in the Dirac\textendash von Neumann axioms
for textbook quantum theory (Dirac 1930, von Neumann 1932)\nocite{Dirac:1930pofm,vonNeumann:1932mgdq}
over precisely which sorts of processes count as measurements. The
category problem refers to the separate problem of accounting for
the larger category of non-measurement phenomena that seem to be happening
in the world around us.\footnote{Despite what one reads in some textbooks on quantum mechanics (for
example, Shankar 1994, Chapter 6)\nocite{Shankar:1994pqm}, one cannot
get around the category problem by appealing to expectation values,
because the only expectation values provided by the Dirac\textendash von
Neumann axioms are averages of numerical \emph{measurement outcomes}
statistically weighted by \emph{measurement-outcome} probabilities.
In the absence of a rigorous argument to the contrary, these expectation
values are categorically narrower than ensemble averages or time averages
of phenomena simply \emph{happening} in the world around us.} 

Based on this reasoning, one can arguably view any quantum system
as an indivisible stochastic process in disguise, leading to what
one could naturally call the \emph{indivisible interpretation of quantum theory},
or just \emph{indivisible quantum theory} for short. Going in the
other logical direction, one can also represent indivisible stochastic
processes as quantum systems in the usual Hilbert-space formalism,
which may open up new applications for Hilbert-space methods.

As a consequence, there exists a \emph{stochastic\textendash quantum correspondence}
between indivisible stochastic systems and quantum systems (Barandes 2025)\nocite{Barandes:2025tsqc}.
This correspondence has the potential to deflate and demystify many
of the exotic features that are usually associated with quantum theory,
with superpositions no longer describing a literal smearing of physical
configurations, and with measurements demoted to just an ordinary
kind of stochastic interaction. Moreover, the correspondence gives
a first-principles way to understand not only why the time evolution
of closed quantum systems is linear, but why it is unitary.\footnote{The conditions of linearity and probability conservation alone are
not sufficient to pick out unitary time evolution. Completely positive
trace-preserving (CPTP) maps, also called quantum channels, need not
be unitary, but are linear and conserve probability. Even adding on
the condition of logical invertibility is not enough to imply unitarity,
because there exist non-unitary quantum channels that are invertible,
such as for a quantum system with a $2\times2$ density matrix $\densitymatrix$
evolving according to the bit-flip channel $\densitymatrix\mapsto{\left(1-p\right)}\densitymatrix+p\sigma_{x}\densitymatrix\sigma_{x}$,
where $p$ is a probability not equal to $1/2$ and where $\sigma_{x}$
is the first Pauli matrix (Chuang 2014)\nocite{Chuang:2014qec}.}

Finally, by replacing the Dirac\textendash von Neumann axioms of textbook
quantum theory\textemdash and their internal ambiguities\textemdash with
a simpler, more internally consistent, and more physically transparent
set of axioms based on configuration spaces and stochastic laws, one
opens up the possibility of new applications and generalizations of
quantum theory. This paper will conclude by describing some of these
downstream implications, as well as questions surrounding causation,
locality, and Bell's theorem.

The goals of this paper are to review the theory of indivisible stochastic
processes, the stochastic\textendash quantum correspondence, and indivisible
quantum theory, as well as to describe the ramifications of the stochastic\textendash quantum
correspondence for dynamical symmetries and for formal enlargements
or dilations of a system's Hilbert space, including several new results.
Along the way, this paper will identify two general forms of gauge
invariance, the first of which has not yet been described in the research
literature.

Attempts to reformulate or reconstruct quantum theory in terms of
stochastic processes, without a fundamental physical role for wave
functions or Hilbert spaces, are far from new. As Everett pointed
out in the unpublished 1956 version of his dissertation, which eventually
led to the \textquoteleft many worlds\textquoteright{} interpretation,
Bopp was already developing a ``stochastic process interpretation''
of quantum theory starting in the 1940s (Everett 1956, pp. 114\textendash 115;
Bopp 1947, 1952, 1953)\nocite{Everett:1956ttotuwf,Bopp:1947qsuk,Bopp:1952efdqbsdk,Bopp:1953sudgdqde}.
Fenyes worked on a similar interpretation in the 1950s (Fenyes 1952)\nocite{Fenyes:1952ewbuidq},
but the most well-known stochastic-process formulation was due to
Nelson's work in the 1960s through the 1980s, and is known today as
Nelsonian stochastic mechanics (Nelson 1966, 1985)\nocite{Nelson:1966dtobm,Nelson:1985qf}.
One can find later examples in the research literature as well (Aaronson
2005; Strocchi 2008, Chapter 6; Frasca 2012; de Oliveira 2025)\nocite{Aaronson:2005qcahv,Strocchi:2008aittmsoqm,Frasca:2012qmitsroasp,DeOliveira:2025csroqm}.
These stochastic approaches were all based on adaptations of methods
like Brownian motion, and so they all assumed Markovian dynamical
laws. This constraint limited the ability of these approaches to capture
the full behavior of quantum systems without becoming very complicated,
and also made it difficult to generalize them beyond systems of fixed
numbers of finitely many non-relativistic particles.

These stochastic approaches are not to be confused with spontaneous-collapse
theories, which treat the wave function or density matrix as a primary
physical ingredient that reduces or collapses probabilistically with
time (Ghirardi, Remini, Weber 1986; Bassi, Ghirardi 2003)\nocite{GhirardiRiminiWeber:1986udmms,BassiGhirardi:2003drm}.
Frameworks forming an entirely separate class, known as general(ized)
probabilistic theories (GPTs), take a thoroughly instrumentalist perspective,
and are phrased in terms of formal, external agents acting on probabilistic
systems (Plavala 2023)\nocite{Plavala:2023gptai}.

In the existing research literature on quantum theory, non-Markovianity
is almost exclusively studied as a phenomenological approximation
to the observable features of \emph{open} quantum systems, for which
deviations from Markovian time evolution arises from interactions
with the environment and associated feedback effects (Breuer, Pettrucione
2002; Milz, Modi 2021)\nocite{BreuerPetruccione:2002toqs,MilzModi:2021qspaqnp}.
As an important exception, Glick and Adami showed that a form of non-Markovianity
is inherent and experimentally accessible in the dynamics of \emph{closed}
quantum systems as well, and is deeply connected with entanglement
(Glick, Adami 2020)\nocite{GlickAdami:2020manmqm}. Glick and Adami
therefore established that a form of non-Markovianity is a fundamental
feature of closed quantum systems, whether one likes it or not. One
can regard the present work as an argument for taking the intrinsic
non-Markovianity of closed quantum systems seriously as a starting
point for \emph{reformulating} quantum theory, in part as a means
toward understanding what quantum theory is really about. Although
the specific definition of non-Markovianity employed by Glick and
Adami does not coincide with \emph{indivisibility}, Glick and Adami
also introduced a notion of Markovianity that coincides with \emph{divisibility},
which refers to a breakdown in indivisibility. As a result, indivisible
quantum theory makes a concrete observational prediction that the
violations of Markovianity identified by Glick and Adami should show
up in suitably arranged experiments, although other interpretations
of quantum theory may make the same prediction as well.

There are very few previous examples of non-Markovian stochastic processes
being used to reformulate\emph{ }quantum theory (Skorobogatov, Svertilov
1998; Gillespie 2000; Gillespie, Alltop, Martin 2001; Dennis 2010)\nocite{SkorobogatovSvertilov:1998qmcbfaansp,Gillespie:2000nsp,GillespieAlltopMartin:2001mqmpaacsp,Dennis:2010nmbvmp}.
Skorobogatov and Svertilov, for example, studied simple, two-level
systems, and showed that with appropriately chosen memory kernels,
one could get adequate predictions, but the paper did not construct
a general theory or framework. Gillespie's approach was more similar
to the one presented here, but focused on defining specific non-Markovian
\emph{realizers}, in the terminology of the present work, rather
than generalizing to indivisible stochastic processes, which had not
been discovered by that point.

Considering the mathematical simplicity of the stochastic\textendash quantum
correspondence between indivisible stochastic processes and quantum
systems, it is surprising that it has apparently not shown up in the
research literature before. To the author's knowledge, the only previous
example that bears a suggestive resemblance to the approach taken
in this paper, at least at the level of some of its equations, is
an unpublished draft by Wetterich (2009)\nocite{Wetterich:2009qrocse}.\footnote{The author thanks Logan McCarty for finding this reference.}
Although that reference argues that some stochastic processes can
be modeled using a formalism similar to that of quantum theory, it
does not establish that the resulting Hilbert-space representation
is fully general. Nor does it attempt to show that the correspondence
is bidirectional, so that quantum systems can be modeled by stochastic
processes in configuration spaces.

Section~\ref{sec:Stochastic-Processes} will review the theory of
indivisible stochastic processes. Section~\ref{sec:Review-of-the-Stochastic-Quantum-Correspondence}
will lay out the stochastic\textendash quantum correspondence, including
the dictionary that ultimately connects indivisible stochastic processes
with quantum systems, as well as identify an important new class of
gauge transformations that have not yet been described in the research
literature. Section~\ref{sec:Further-Implications} will describe
further implications of the stochastic\textendash quantum correspondence,
focusing on dynamical symmetries and Hilbert-space dilations. Section~\ref{sec:Discussion-and-Future-Work}
will conclude with a summary and a discussion of future work.

\section{Stochastic Processes\label{sec:Stochastic-Processes}}

\subsection{Textbook Stochastic Processes\label{subsec:Textbook-Stochastic-Processes}}

According to widely used textbooks (Rosenblatt 1962, Parzen 1962,
Doob 1990, Ross 1995)\nocite{Rosenblatt:1962rp,Parzen:1962sp,Doob:1990sp,Ross:1995sp},
the most general kind of \emph{stochastic process} requires only
a \emph{sample space}, a \emph{probability distribution}, and one
or more \emph{time-dependent random variables}, meaning  time-indexed
families of functions from the sample space to the real numbers. However,
stochastic processes defined in this narrow way lack an ingredient
that plays the role of a dynamical law. This paper will be concerned
with a slightly modified construction that includes the notion of
a dynamical law.

As an important case, one can naturally model classical physical systems
as stochastic processes, ultimately with dynamical laws that may be
probabilistic in only a trivial sense. In defining a stochastic process
to serve as such a model, one can take the sample space to be the
system's \emph{configuration space} $\configspace$,\footnote{For Newtonian systems, one could instead take the sample space to
be the system's \emph{phase space}, provided that one writes the
dynamical laws in a different way.} which is a \emph{fixed} ingredient of the model, meaning that it
remains the same for every physical run or instantiation of the model.
Examples would include the discrete configurations of a system of
finitely many digital bits, or the continuous set of possible arrangements
of a collection of particles in three-dimensional physical space.

For each physical instantiation of the physical system to be modeled,
one then has a \emph{standalone probability distribution} $p{\left(i,t\right)}$,
where $i$ denotes a configuration of the system, $t$ is the time,
and $p{\left(i,t\right)}$ is the probability or probability density
for the system to be in its $i$th configuration at the time $t$.
The system's standalone probability distribution is a \emph{contingent}
feature of the model, meaning that it can be different each time the
model is run or instantiated. As a probability distribution, $p{\left(i,t\right)}$
satisfies the usual conditions 
\begin{equation}
0\leq p{\left(i,t\right)}\leq1,\quad\sum_{i}p{\left(i,t\right)}=1,\label{eq:StandaloneProbabilityDistributionKolmogorovConditions}
\end{equation}
 where the discrete summation would be replaced with an integration
in the case of a system with a continuous configuration space.

Time-dependent maps $A{\left(t\right)}$ from the system's configuration
space $\configspace$ to the real numbers are then \emph{random variables},
with individual values or magnitudes $a{\left(i,t\right)}$ that depend
both on the system's configuration $i$ as well as explicitly on the
time $t$. Random variables can also inherit an implicit time-dependence
from the contingent standalone probability distribution $p{\left(i,t\right)}$,
which probabilistically determines the system's configuration $i$
and therefore the random variable's value $a{\left(i,t\right)}$.
The statistical average or expectation value of a random variable
$A{\left(t\right)}$ is then defined as the probability-weighted sum
\begin{equation}
\expectval{A{\left(t\right)}}\defeq\sum_{i}a{\left(i,t\right)}p{\left(i,t\right)},\label{eq:DefExpectationValueRandomVariable}
\end{equation}
 where the two forms of time-dependence are manifest on the right-hand
side\textemdash explicit in $a{\left(i,t\right)}$ and implicit in
$p{\left(i,t\right)}$\textemdash and, again, the discrete summation
would be replaced with an integration if the system has a continuous
configuration space.

As explained already, the configuration space $\configspace$ is one
of the \emph{fixed} ingredients of the model, and provides the model
with its kinematics, meaning its elementary physical or \textquoteleft ontological\textquoteright{}
content. This configuration space should be selected on a case-by-case
basis for whatever kind of system one is interested in modeling, and
this high degree of generality should be seen as an intended virtue
of this paper's overall approach. For a system of discrete bits, the
configuration space would consist of bivalent degrees of freedom,
but one could also, in principle, handle systems of particles, for
which the configuration space would consist of particle arrangements
(with or without some level of coarse-graining), or one could handle
systems of fields, for which the configuration space would consist
of patterns of field intensities and polarizations.

Meanwhile, the standalone probability distribution $p{\left(i,t\right)}$
is a \emph{contingent} ingredient. As will follow from the discussion
ahead, this standalone probability distribution will generically blend
together information about the model's initial conditions along with
some amount of \textquoteleft nomological\textquoteright{} content
from the model's dynamical laws. In principle at least, $p{\left(i,t\right)}$
may involve some \textquoteleft epistemic\textquoteright{} content
as well, in the sense of subjective credences.

A textbook stochastic process is, moreover, generally assumed to have
a very intricate set of elementary dynamical laws, or \textquoteleft nomological\textquoteright{}
content, consisting of an infinite hierarchy or \emph{Kolmogorov tower}
of first-order, second-order, third-order, and higher-order conditional
probabilities, conditioned on a sequence of times $t_{1},t_{2},t_{3},\dots$
increasingly into the past, and satisfying $t>t_{1}>t_{2}>t_{3}>\cdots$:\footnote{This nomological structure is emphasized, for example, in the work
of Gillespie and collaborators (Gillespie 1996, Section IV; Gillespie
1998; Gillespie, Alltop, Martin 1999; Gillespie 2000) \nocite{Gillespie:1996tmlafpe,Gillespie:1998dtsoasp,GillespieAlltopMartin:1999codtsoaspbdtgajp66551,Gillespie:2000nsp}.} 
\begin{equation}
{\left.\begin{aligned}p{\left(i,t\given j_{1},t_{1}\right)} & \quad{\left[\textrm{first-order}\right]},\\
p{\left(i,t\given j_{1},t_{1};j_{2},t_{2}\right)} & \quad{\left[\textrm{second-order}\right]},\\
p{\left(i,t\given j_{1},t_{1};j_{2},t_{2};j_{3},t_{3}\right)}, & \quad{\left[\textrm{third-order}\right]},\\
 & \vdots
\end{aligned}
\right\} }\label{eq:TowerConditionalProbabilities}
\end{equation}
 Here vertical separators $\vert$ should be read as \textquoteleft given\textquoteright{}
and semicolons $;$ should be read as the \textquoteleft and\textquoteright{}
operator. In each of these conditional probabilities, the single time
$t$ to the left of the vertical separator will be called a \emph{target time},
and the times $t_{1},t_{2},t_{3},\dots$ to the right of the vertical
separator will be called \emph{conditioning times}.

Essentially, at the conditioning times, one specifies \textquoteleft initial
conditions\textquoteright{} that ultimately lead to a probabilistic
prediction for the system's configuration at the target time. To make
clear this important dynamical role played by the Kolmogorov tower
\eqref{eq:TowerConditionalProbabilities}, these conditional probabilities
will also be called \emph{transition probabilities}. It is somewhat
more intuitive to think of them as aleatory (\textquoteleft objective
chance\textquoteright ) probabilities, rather than as epistemic (\textquoteleft subjective
credence\textquoteright ) probabilities.

Unlike the \emph{contingent} standalone probability distribution $p{\left(i,t\right)}$,
one might wish to view the transition probabilities \eqref{eq:TowerConditionalProbabilities}
as \emph{fixed} rules built into the laws of the model, and thus as
features that are not supposed to change in value for each run or
instantiation of the model. However, by combining them with the contingent
standalone probability distribution, one can construct contingent
\emph{joint} probabilities of greater and greater complexity: 
\begin{equation}
{\left.\begin{aligned}p{\left(i,t;j_{1},t_{1}\right)}\defeq p{\left(i,t\given j_{1},t_{1}\right)}p{\left(j_{1},t_{1}\right)} & \quad{\left[\textrm{two-way}\right]},\\
p{\left(i,t;j_{1},t_{1};j_{2},t_{2}\right)}\defeq p{\left(i,t\given j_{1},t_{1};j_{2},t_{2}\right)}p{\left(j_{1},t_{1};j_{2},t_{2}\right)} & \quad{\left[\textrm{three-way}\right]},\\
 & \vdots
\end{aligned}
\right\} }\label{eq:TowerJointProbabilities}
\end{equation}
 Said somewhat more precisely, these joint probabilities mix together
contingent-epistemic and fixed-nomological data\textemdash they are
not fixed and purely nomological, in contrast with the transition
probabilities \eqref{eq:TowerConditionalProbabilities}. The joint
probabilities are, in turn, required to satisfy an intricate set of
marginalization consistency conditions known as Chapman\textendash Kolmogorov
equations: 
\begin{equation}
\sum_{j_{k}}p{\left(\dots;j_{k},t_{k};\dots\right)}=p{\left(\dots{\left[\textrm{no }j_{k}\right]}\dots\right)}.\label{eq:ChapmanKolmogorovEquations}
\end{equation}
 That is, 
\begin{equation}
{\left.\begin{aligned}\sum_{i}p{\left(i,t;j_{1},t_{1}\right)} & =p{\left(j_{1},t_{1}\right)},\\
\sum_{j_{1}}p{\left(i,t;j_{1},t_{1}\right)} & =p{\left(i,t\right)},\\
\sum_{i}p{\left(i,t;j_{1},t_{1};j_{2},t_{2}\right)} & =p{\left(j_{1},t_{1};j_{2},t_{2}\right)},\\
\sum_{j_{1}}p{\left(i,t;j_{1},t_{1};j_{2},t_{2}\right)} & =p{\left(i,t;j_{2},t_{2}\right)},
\end{aligned}
\right\} }\label{eq:ChapmanKolmogorovEquationsExplicit}
\end{equation}
 and so forth. There are many other consistency conditions that must
be satisfied as well (Gillespie, Alltop, Martin 1999)\nocite{GillespieAlltopMartin:1999codtsoaspbdtgajp66551}.

As mentioned earlier, the framework described here is general enough
to accommodate classical, deterministic systems. For example, a Newtonian
system based on deterministic, second-order equations of motion is
equivalent to a stochastic process whose nomological content consists
of only second-order conditional probabilities, with values equal
to $1$ or $0$ (or, more properly delta functions), and with times
$t,t_{1},t_{2}$ separated by vanishingly small durations.

A stochastic process is said to be \emph{Markovian} if all its second-
and higher-order transition probabilities $p{\left(i,t\given j_{1},t_{1};j_{2},t_{2},\dots\right)}$
exist and are equal in value to the \emph{first-order} transition
probabilities $p{\left(i,t\given j_{1},t_{1}\right)}$ for the time
$t_{1}$ closest to $t$. That is, for a Markov process, the whole
Kolmogorov tower \eqref{eq:TowerConditionalProbabilities} of transition
probabilities is included in the model, but the latest conditioning
time $t_{1}$ always \textquoteleft screens off\textquoteright{} all
the earlier conditioning times $t_{2},\dots<t_{1}$, thereby trivializing
all second- and higher-order transitions probabilities. Crucially,
all those second- and higher-order transition probabilities still
\emph{exist} in the laws\textemdash the point is just that their numerical
values are all determined by first-order transition probabilities.
It follows immediately from the Chapman\textendash Kolmogorov equations
\eqref{eq:ChapmanKolmogorovEquations} that the first-order transition
probabilities concatenate or compose over time, in the sense that
\begin{equation}
p{\left(i,t\right)}=\sum_{j_{1},j_{2},\dots}p{\left(i,t\given j_{1},t_{1}\right)}p{\left(j_{1},t_{1}\given j_{2},t_{2}\right)}\cdots p{\left(j_{n-1},t_{n-1}\given j_{n},t_{n}\right)}p{\left(j_{n},t_{n}\right)}.\label{eq:MarkovProcessDivisibility}
\end{equation}
 Notice, in particular, that the time-evolution laws from $t_{n}$
to $t$ decompose or \emph{\textquoteleft divide\textquoteright{}}
into time-evolution laws for all the intermediate time steps. The
fact that Markovianity implies divisibility will be important to keep
in mind, because it follows then, as the contrapositive implication,
that indivisibility will entail non-Markovianity.

One can introduce some helpful notation to make this divisibility
of Markovian dynamics more manifest. Observe that each first-order
transition probability appearing in \eqref{eq:MarkovProcessDivisibility}
is labeled by a configuration $i$ at a target time $t$, and by another
configuration $j$ at a conditioning time $t^{\prime}$. One can therefore
organize these first-order transition probabilities into a \emph{transition matrix}
$\stochasticmatrix{\left(t\from t^{\prime}\right)}$ with individual
entries defined according to 
\begin{equation}
\stochasticmatrix_{ij}{\left(t\from t^{\prime}\right)}\defeq p{\left(i,t\given j,t^{\prime}\right)}.\label{eq:DefStochasticMatrixAsConditionals}
\end{equation}
 Because the entries of $\stochasticmatrix{\left(t\from t^{\prime}\right)}$
are all conditional probabilities, they are all non-negative, and
each column sums to $1$, meaning that $\stochasticmatrix{\left(t\from t^{\prime}\right)}$
is a \emph{(column) stochastic matrix}: 
\begin{equation}
\stochasticmatrix_{ij}{\left(t\from t^{\prime}\right)}\geq0,\quad\sum_{i}\stochasticmatrix_{ij}{\left(t\from t^{\prime}\right)}=1.\label{eq:TransitionMatrixIsColumnStochastic}
\end{equation}
 Introducing a column vector $p{\left(t\right)}$ with individual
entries given by 
\begin{equation}
p_{i}{\left(t\right)}\defeq p{\left(i,t\right)},\label{eq:DefProbabilityColumnVec}
\end{equation}
 one can write the composition equation \eqref{eq:MarkovProcessDivisibility}
as 
\begin{equation}
p_{i}{\left(t\right)}=\sum_{j_{1},j_{2},\dots}\stochasticmatrix_{ij_{1}}{\left(t\from t_{1}\right)}\stochasticmatrix_{j_{1}j_{2}}{\left(t_{1}\from t_{2}\right)}\cdots\stochasticmatrix_{j_{n-1}j_{n}}{\left(t_{n-1}\from t_{n}\right)}p_{j_{n}}{\left(t_{n}\right)},\label{eq:MarkovProcessDivisibilityMatrixExplicit}
\end{equation}
 or, more compactly, as 
\begin{equation}
p{\left(t\right)}=\stochasticmatrix{\left(t\from t_{1}\right)}\stochasticmatrix{\left(t_{1}\from t_{2}\right)}\cdots\stochasticmatrix{\left(t_{n-1}\from t_{n}\right)}p{\left(t_{n}\right)}.\label{eq:MarkovProcessDivisibilityMatrix}
\end{equation}
 Treating the standalone probabilities here as contingent and arbitrary,
it follows that the transition matrix $\stochasticmatrix{\left(t\from t_{n}\right)}$
from $t_{n}$ all the way to $t$ \emph{divides} into a product of
transition matrices for all the subintervals: 
\begin{equation}
\stochasticmatrix{\left(t\from t_{n}\right)}=\stochasticmatrix{\left(t\from t_{1}\right)}\stochasticmatrix{\left(t_{1}\from t_{2}\right)}\cdots\stochasticmatrix{\left(t_{n-1}\from t_{n}\right)}.\label{eq:MarkovProcessDivisibilityAbstract}
\end{equation}
 In particular, given any two times $t$ and $t_{0}$, with $t>t_{0}$,
then for any intermediate time $t^{\prime}$ between them, the system's
transition matrices satisfy the \emph{divisibility condition} 
\begin{equation}
\stochasticmatrix{\left(t\from t_{0}\right)}=\stochasticmatrix{\left(t\from t^{\prime}\right)}\stochasticmatrix{\left(t^{\prime}\from t_{0}\right)}.\label{eq:DivisibilityCondition}
\end{equation}

The simplest example of a Markov process is a \emph{discrete-time, time-homogeneous Markov chain}
based on a finite time scale $\delta t$, for which the time-dependent
transition matrix \eqref{eq:DefStochasticMatrixAsConditionals} at
any integer number $n\geq1$ of steps of duration $\delta t$ from
an initial time $0$ can be expressed as $n$ powers of a stochastic
transition matrix $\stochasticmatrix$ that is constant in time: 
\begin{equation}
\stochasticmatrix{\left(n\,\delta t\from0\right)}=\stochasticmatrix^{n}.\label{eq:MarkovChainTransitionMatrixAsPowers}
\end{equation}

A textbook stochastic process that fails to be Markovian is said to
be \emph{non-Markovian}, and, in some cases, can be regarded as exhibiting
\emph{memory} effects. Memory here refers to the fact that conditioning
on more and more previous times changes the values of the conditional
probabilities: 
\begin{equation}
{\left.\begin{aligned}p{\left(i,t\given j_{1},t_{1}\right)} & \ne p{\left(i,t\given j_{1},t_{1};j_{2},t_{2}\right)}\\
 & \ne p{\left(i,t\given j_{1},t_{1};j_{2},t_{2};j_{3},t_{3}\right)}\\
 & \cdots.
\end{aligned}
\right\} }\label{eq:NonMarkovianMemoryProcessUnequalConditionalProbabilities}
\end{equation}
 The Chapman\textendash Kolmogorov equations \eqref{eq:ChapmanKolmogorovEquations}
then yield equations that are more complicated than the Markovian
case \eqref{eq:MarkovProcessDivisibility}, such as 
\begin{equation}
p{\left(i,t\right)}=\sum_{j_{1},j_{2},j_{3},\dots}p{\left(i,t\given j_{1},t_{1};j_{2},t_{2};j_{3},t_{3}\right)}p{\left(j_{1},t_{1}\given j_{2},t_{2};j_{3},t_{3}\right)}p{\left(j_{2},t_{2}\given j_{3},t_{3}\right)}p{\left(j_{3},t_{3}\right)}.\label{eq:NonMarkovianProcessDivisibility}
\end{equation}
 The simplest kind of non-Markovian stochastic process has an infinite
Kolmogorov tower \eqref{eq:TowerConditionalProbabilities} that trivializes
after $n$ orders, for some finite integer $n$. Such a stochastic
process is said to be \emph{$n$th-order non-Markovian} (or \emph{$n$th-order Markovian}).
A Markov process corresponds to the case $n=1$. Newtonian mechanics,
viewed as a deterministic version of a stochastic process, is second-order
non-Markovian, because Newton's second law is a second-order differential
equation, so it requires specifying positions at the present time
and at a time infinitesimally in the past\textemdash although, in
practice, one instead specifies positions and \textquoteleft instantaneous\textquoteright{}
velocities at just the present time.

\subsection{Indivisible Stochastic Processes\label{subsec:Indivisible-Stochastic-Processes}}

An \emph{indivisible stochastic process} generalizes a textbook stochastic
process in a very simple way (Barandes 2025)\nocite{Barandes:2025tsqc}.
Rather than laws consisting of an infinite Kolmogorov tower of conditional
probabilities of arbitrarily high order, the laws of an indivisible
stochastic process contain only first-order transition probabilities
connecting target times $t$ with conditioning times $t_{0}$, making
up a transition matrix $\stochasticmatrix{\left(t\from t_{0}\right)}$
with individual entries 
\begin{equation}
\stochasticmatrix_{ij}{\left(t\from t_{0}\right)}\defeq p{\left(i,t\given j,t_{0}\right)}.\label{eq:IndivisibleStochasticProcessConditionalProbabilities}
\end{equation}
 The only available Chapman\textendash Kolmogorov equations \eqref{eq:ChapmanKolmogorovEquations}
take the simple form 
\begin{equation}
p{\left(i,t\right)}=\sum_{j}p{\left(i,t\given j,t_{0}\right)}p{\left(j,t_{0}\right)},\label{eq:IndivisibleStochasticProcessChapmanKolmogorovEquationsLawOfTotalProbability}
\end{equation}
 which is just the law of total probability. Equivalently, in matrix
notation, 
\begin{equation}
p{\left(t\right)}=\stochasticmatrix{\left(t\from t_{0}\right)}p{\left(t_{0}\right)}.\label{eq:IndivisibleStochasticProcessChapmanKolmogorovEquationsMatrixForm}
\end{equation}
 For continuity, the transition matrix $\stochasticmatrix{\left(t\from t_{0}\right)}$
will typically be assumed to approach the identity matrix $\idmatrix$
in the limit $t\to t_{0}$: 
\begin{equation}
\lim_{t\to t_{0}}\stochasticmatrix{\left(t\from t_{0}\right)}=\idmatrix.\label{eq:ContinuityConditionInitialTimeTransitionMatrix}
\end{equation}

Importantly, notice that the law of total probability \eqref{eq:IndivisibleStochasticProcessChapmanKolmogorovEquationsLawOfTotalProbability}
is \emph{linear}, in the sense that it establishes a linear relationship
between the system's standalone probabilities at $t_{0}$ and the
system's standalone probabilities at $t$. This linearity will turn
out to underwrite the linearity of time evolution for closed quantum
systems.

Note that no assumption is made here that the transition probabilities
$p{\left(i,t\given j,t_{0}\right)}$ exist as part of the laws for
all real-valued choices of the conditioning time $t_{0}$. The conditioning
times $t_{0}$ that are allowed in these nomological transition probabilities
are called \emph{division events} for the given system, and, without
any real loss of generality, will be assumed to include an \textquoteleft initial\textquoteright{}
time $0$. (More will be said below about the status of conditioning
times \emph{outside} of transition probabilities.) Division events
will play an important role in this framework, so it will be worthwhile
to dwell for a moment on their properties and significance.

Division events are not global properties of the whole universe, but
are system-centric, just like various other kinds of spontaneous time-translation-breaking
in physics. In practice, a system may have multiple exact division
events, or they may be generated to an extremely good approximation
through interactions with other systems, after marginalizing over
those other systems (Barandes 2025)\nocite{Barandes:2025tsqc}.

The target time $t$, by contrast, can be treated as a free and continuous
variable, so there is no fundamental assumption of discreteness of
time. Also, in particular, no assumption is made that $t>t_{0}$.
One can choose $t<t_{0}$ as well. An indivisible stochastic process
does not, therefore, need to violate logical time-reversal invariance
in any fundamental way, although time-reversal invariance, like time-translation
invariance, may be broken spontaneously for a given system through
interactions with other systems.

Crucially, an indivisible stochastic process, as befits its name,
will not generally obey a divisibility condition like \eqref{eq:DivisibilityCondition}.
Given transition matrices $\stochasticmatrix{\left(t\from t_{0}\right)}$
and $\stochasticmatrix{\left(t^{\prime}\from t_{0}\right)}$, where
$t^{\prime}$ lies in the interval between $t_{0}$ and $t$, it might
seem reasonable to try to define an intermediate transition matrix
$\tilde{\stochasticmatrix}{\left(t\from t^{\prime}\right)}$ from
$t^{\prime}$ to $t$ according to 
\begin{equation}
\tilde{\stochasticmatrix}{\left(t\from t^{\prime}\right)}\defeq\stochasticmatrix{\left(t\from t_{0}\right)}\stochasticmatrix^{-1}{\left(t^{\prime}\from t_{0}\right)},\label{eq:DefCandidateRelativeTransitionMatrix}
\end{equation}
 at least if $\stochasticmatrix{\left(t^{\prime}\from t_{0}\right)}$
is invertible. By construction, it would then seem to follow that
the system obeys a divisibility condition akin to \eqref{eq:DivisibilityCondition}:
\begin{equation}
\stochasticmatrix{\left(t\from t_{0}\right)}=\tilde{\stochasticmatrix}{\left(t\from t^{\prime}\right)}\stochasticmatrix{\left(t^{\prime}\from t_{0}\right)}.\label{eq:CandidateRelativeTransitionMatrixInDivisibilityEquation}
\end{equation}
 However, it turns out that a matrix $\tilde{\stochasticmatrix}{\left(t\from t^{\prime}\right)}$
defined according to \eqref{eq:DefCandidateRelativeTransitionMatrix}
will generically fail to be a column stochastic matrix, and, indeed,
will typically have negative entries, and so will form a so-called
\emph{pseudo-stochastic matrix} (Chru\'{s}ci\'{n}ski, Man'ko, Marmo,
Ventriglia 2015)\nocite{ChruscinskiMankoMarmoVentriglia:2015opmapm}.
The reason is that the inverse of a stochastic matrix can only itself
be a stochastic matrix if both matrices are \emph{permutation matrices},
meaning that they consist solely of $0$s and $1$s, and therefore
do not contain nontrivial probabilities.\footnote{Proof: Let $X$ and $Y$ be $N\times N$ matrices with only non-negative
entries and with $Y=X^{-1}$, so that $XY=\idmatrix$. Then, in particular,
the first row of $X$ must be orthogonal to the second-through-$N$th
columns of $Y$. Because, by assumption, $Y$ is invertible, the columns
of $Y$ must all be linearly independent, so the first row of $X$
must be orthogonal to the ${\left(N-1\right)}$-dimensional subspace
spanned by the second-through-$N$th columns of $Y$. Because the
entries of $X$ and $Y$ are all non-negative by assumption, the only
way that this orthogonality condition can hold is if precisely one
of the entries in the first row of $X$ is nonzero, with a $0$ in
the corresponding entry in each of the second-through-$N$th columns
of $Y$. Repeating this argument for the other rows of $X$, one sees
that $X$ can only have a single nonzero entry in each row. If $X$
is a stochastic matrix, then each of these nonzero entries must be
the number $1$, so $X$ must be a permutation matrix. Because the
inverse of a permutation matrix is again a permutation matrix, it
follows that $Y$ must likewise be a permutation matrix.~QED} Then even when such a matrix $\tilde{\stochasticmatrix}{\left(t\from t^{\prime}\right)}$
may accidentally happen to contain only non-negative entries, it should
not be trusted to give the correct dynamical rule for the system's
evolution, at least unless $t^{\prime}$ is a division event. Indeed,
if the system's time evolution from $t_{0}$ to $t$ is \emph{unistochastic},
a term to be defined shortly, then intermediate times $t^{\prime}$
between $t_{0}$ and $t$ \emph{provably} cannot be division events.

Given just the minimalist ingredients that define a given indivisible
stochastic process, there will generically exist a large or infinite
number of ways of choosing a complete Kolmogorov tower \eqref{eq:TowerConditionalProbabilities}
consistent with those ingredients. Any such Kolmogorov tower that
happens to match the statistical patterns in the system's underlying
behavior will be a contingent fact. The conditional probabilities
in any such Kolmogorov tower can be second- or higher order in conditioning
times. Moreover, the conditioning times appearing in a Kolmogorov
tower need not be limited to division events, in contrast to the restriction
of conditioning times to division events in the system's nomological
transition probabilities.

Each such choice of Kolmogorov tower is called a non-Markovian \emph{realizer}.\footnote{The author thanks Alexander Meehan for suggesting this terminology.}
A given indivisible stochastic process will therefore encompass a
whole equivalence class of non-Markovian realizers. Because only the
minimalist ingredients specified by the indivisible stochastic process
will end up being connected to the empirical predictions of quantum
theory, the specific non-Markovian realizer is not only contingent,
but is potentially unknowable, and perhaps meaningless. Indeed, from
a more fundamental perspective, the system ultimately takes only one
trajectory, and if one knew that trajectory in detail, then probabilities
would not strictly be needed in the first place.

Rather surprisingly, one can model quantum systems as indivisible
stochastic processes for suitable choices of the first-order conditional
probabilities \eqref{eq:IndivisibleStochasticProcessConditionalProbabilities}
that make up their laws, as will be explained in the next section.

\section{Review of the Stochastic\textendash Quantum Correspondence\label{sec:Review-of-the-Stochastic-Quantum-Correspondence}}

\subsection{The Time-Evolution Operator\label{subsec:The-Time-Evolution-Operator}}

To see how this construction works in the finite-dimensional case,
consider an indivisible stochastic process with $N$ total configurations
$i=1,\dots,N$ making up the system's configuration space $\configspace$.
Letting $0$ denote a suitable choice of initial division event, define
the system's time-dependent, $N\times N$ transition matrix as usual
according to \eqref{eq:IndivisibleStochasticProcessConditionalProbabilities}:
\begin{equation}
\stochasticmatrix_{ij}{\left(t\from0\right)}\defeq p{\left(i,t\given j,0\right)}.\label{eq:IndivisibleStochasticTransitionMatrix}
\end{equation}
 Then solve the inequality $\stochasticmatrix_{ij}{\left(t\from0\right)}\geq0$
appearing in \eqref{eq:TransitionMatrixIsColumnStochastic} by introducing
a \textquoteleft potential\textquoteright{} consisting of complex-valued
matrix elements $\dynop_{ij}{\left(t\from0\right)}$ related to $\stochasticmatrix_{ij}{\left(t\from0\right)}$
according to the modulus-squaring operation: 
\begin{equation}
\stochasticmatrix_{ij}{\left(t\from0\right)}=\verts{\dynop_{ij}{\left(t\from0\right)}}^{2}.\label{eq:StochasticMatrixEntryFromAbsValSquare}
\end{equation}
 Note that this formula is not a postulate, but an identity, and that
the potential matrix $\dynop{\left(t\from0\right)}$, which will be
called a \emph{time-evolution operator} in what follows, is not unique.\footnote{No assumption is made at this point that the time-evolution operator
$\dynop{\left(t\from0\right)}$ is unitary. Also, the terms \textquoteleft matrix\textquoteright{}
and \textquoteleft operator\textquoteright{} will be used synonymously
in this paper.} It follows from the other formula $\sum_{i}\stochasticmatrix_{ij}{\left(t\from0\right)}=1$
in \eqref{eq:TransitionMatrixIsColumnStochastic} that the time-evolution
operator $\dynop{\left(t\from0\right)}$ satisfies the summation condition
\begin{equation}
\sum^{N}_{i=1}\verts{\dynop_{ij}{\left(t\from0\right)}}^{2}=1.\label{eq:SumTimeEvOpAbsValSqEq1}
\end{equation}

There are several helpful ways to re-express the identity \eqref{eq:StochasticMatrixEntryFromAbsValSquare}.
To begin, one can introduce the \emph{Schur\textendash Hadamard product}
$\hadamardprod$, which is defined for arbitrary $N\times N$ matrices
$X$ and $Y$ as entry-wise multiplication (Schur 1911, Halmos 1958,
Horn 1990)\nocite{Schur:1911bztdbbmuvv,Halmos:1958fvs,Horn:1990thp}:
\begin{equation}
{\left(X\hadamardprod Y\right)}_{ij}\defeq X_{ij}Y_{ij}\quad{\left[\textrm{no sum on repeated indices}\right]}.\label{eq:DefSchurHadamardProduct}
\end{equation}
 One can then regard \eqref{eq:StochasticMatrixEntryFromAbsValSquare}
as expressing the transition matrix $\stochasticmatrix{\left(t\from0\right)}$
as a Schur\textendash Hadamard factorization of the complex-conjugated
potential matrix $\overconj{\dynop{\left(t\from0\right)}}$ with $\dynop{\left(t\from0\right)}$
itself: 
\begin{equation}
\stochasticmatrix{\left(t\from0\right)}={\left|\dynop\left(t\from0\right)\right|}^{\hadamardprod2}\defeq\overconj{\dynop{\left(t\from0\right)}}\hadamardprod\dynop{\left(t\from0\right)}.\label{eq:StochasticMatrixSchurHadamardFactorization}
\end{equation}

Schur\textendash Hadamard products are not widely used in linear algebra,
in part because they are basis-dependent. For the purposes of analyzing
a given indivisible stochastic process, however, this basis-dependence
is not a problem, because the system's configuration space $\configspace$
naturally singles out a specific basis, to be defined momentarily.

\subsection{Schur\textendash Hadamard Gauge Transformations\label{subsec:Schur-Hadamard-Gauge-Transformations}}

To make the nonuniqueness of the potential matrix $\dynop{\left(t\from0\right)}$
more manifest, it will be helpful to introduce an analogy with the
Maxwell theory of classical electromagnetism.\footnote{For pedagogical treatments of classical electromagnetism, see Griffiths
(2023), Zangwill (2012), or Jackson (1998)\nocite{Griffiths:2023ite,Zangwill:2012me,Jackson:1998ce}.}

In classical electromagnetism, the electric and magnetic fields are
physically meaningful quantities, but it is often very convenient
to work instead in terms of scalar and vector potentials, which are
not uniquely defined. All choices for the potentials that yield the
same electric and magnetic fields are said to be related by gauge
transformations. Any one such choice for the potentials is called
a gauge choice, and the scalar and vector potentials themselves are
called gauge potentials or gauge variables.

Making a suitable gauge choice can greatly simplify many calculations,
such as in using Lorenz gauge to compute the electric and magnetic
fields for delayed boundary conditions. Ultimately, however, the theory
does not treat any gauge choice as fundamentally more correct than
any other gauge choice, and all calculations of physical predictions
in classical electromagnetism must conclude with expressions that
are written in terms of gauge-invariant quantities.

To set up the claimed analogy with electromagnetic gauge transformations,
one starts by observing from the basic relationship $\stochasticmatrix_{ij}{\left(t\from0\right)}=\verts{\dynop_{ij}{\left(t\from0\right)}}^{2}$
in \eqref{eq:StochasticMatrixEntryFromAbsValSquare} that the Schur\textendash Hadamard
product \eqref{eq:DefSchurHadamardProduct} of the time-evolution
operator $\dynop{\left(t\from0\right)}$ and a matrix of arbitrary,
time-dependent phases $\exp\parens{i\theta_{ij}\parens t}$ is a transformation
of $\dynop{\left(t\from0\right)}$ with no physical effects, at least
under the assumption that one has made an \textquoteleft ontologically\textquoteright{}
correct choice of configuration space for the underlying indivisible
stochastic process, or if the choice of configuration space is sufficiently
rich that all measurement records can supervene on it. This transformation
then corresponds to a genuine form of gauge invariance: 
\begin{equation}
\dynop{\left(t\from0\right)}\mapsto\dynop{\left(t\from0\right)}\hadamardprod\begin{pmatrix}e^{i\theta_{11}{\left(t\right)}} & e^{i\theta_{12}{\left(t\right)}}\\
e^{i\theta_{21}{\left(t\right)}} & \ddots\\
 &  & e^{i\theta_{NN}{\left(t\right)}}
\end{pmatrix}.\label{eq:DefLocalInTimeSchurHadamardGaugeTransformation}
\end{equation}
 This gauge transformation can be written equivalently at the level
of individual matrix entries as 
\begin{equation}
\dynop_{ij}{\left(t\from0\right)}\mapsto\dynop_{ij}{\left(t\from0\right)}e^{i\theta_{ij}{\left(t\right)}}.\label{eq:DefSchurHadamardGaugeTransformationEntries}
\end{equation}

Although it may seem surprising that these time-dependent changes
of phase have no physical effects, keep in mind the \emph{crucial}
fact that this gauge transformation will end up having downstream
effects on the definitions of various Hilbert-space ingredients ahead,
in just such a way that all empirical predictions will remain unchanged.\footnote{In particular, this form of gauge invariance is \emph{not} equivalent
to changing merely the relative phases of state vectors \emph{alone}.}

This kind of gauge transformation will be called a \emph{Schur\textendash Hadamard gauge transformation},
because it involves the sort of entry-wise multiplication that characterizes
Schur\textendash Hadamard products. To the author's knowledge, this
kind of gauge invariance has not yet been described in the research
literature, and is therefore a new result. It will turn out to play
a key role in the analysis of dynamical symmetries that will be presented
in Subsection~\ref{subsec:Dynamical-Symmetries}, and will be extended
in an interesting way in the context of Hilbert-space dilations in
Subsection~\ref{subsec:Dilations}. A different but equally general
form of gauge invariance will be discussed later on.

\subsection{The Dictionary\label{subsec:The-Dictionary}}

As an important alternative approach to writing the identity $\stochasticmatrix_{ij}{\left(t\from0\right)}=\verts{\dynop_{ij}{\left(t\from0\right)}}^{2}$
in \eqref{eq:StochasticMatrixEntryFromAbsValSquare}, one starts by
introducing a set of $N\times1$ column vectors $e_{1},\dots,e_{N}$,
where $e_{i}$ has a $1$ in its $i$th component and $0$s in all
its other components: 
\begin{equation}
e_{1}\defeq{\left(\begin{smallmatrix}1\\
0\\
\vdots\\
0\\
0
\end{smallmatrix}\right)},\quad\dots,\quad e_{N}\defeq{\left(\begin{smallmatrix}0\\
0\\
\vdots\\
0\\
1
\end{smallmatrix}\right)}.\label{eq:DefConfigurationBasis}
\end{equation}
 That is, $e_{i}$ has components 
\begin{equation}
e_{i,j}=\delta_{ij}.\label{eq:DefConfigurationBasisEntries}
\end{equation}
 It follows that the column vectors $e_{1},\dots,e_{N}$ form an orthonormal
basis for the vector space of all $N\times1$ column vectors, and
$e_{1},\dots,e_{N}$ will be called the system's \emph{configuration basis}.

In particular, 
\begin{equation}
e^{\adj}_{i}e_{j}=\delta_{ij},\quad e_{i}e^{\adj}_{i}=\projector_{i},\label{eq:ConfigurationBasisOrthonormalComplete}
\end{equation}
 where $\delta_{ij}$ is the usual Kronecker delta, 
\begin{equation}
\delta_{ij}\defeq\begin{cases}
1 & \textrm{for }i=j,\\
0 & \textrm{for }i\ne j,
\end{cases}\label{eq:DefKroneckerDelta}
\end{equation}
 and where $\projector_{1},\dots,\projector_{N}$ are an $N$-member
collection of $N\times N$ constant, diagonal projection matrices,
which will be called \emph{configuration projectors}. For each $i=1,\dots,N$,
the configuration projector $\projector_{i}$ consists of a single
$1$ in its $i$th row, $i$th column, and $0$s in all its other
entries. That is, $\projector_{i}$ is defined as 
\begin{equation}
\projector_{i}\defeq\mathrm{diag}\parens{0,\dots,0,\underset{\mathclap{\substack{\uparrow\\
i\textrm{th entry}
}
}}{1},0,\dots,0}=e_{i}e^{\adj}_{i},\label{eq:DefConfigurationProjectorsAsDiagonalMatrices}
\end{equation}
 with individual entries 
\begin{equation}
\projector_{i,jk}=\delta_{ij}\delta_{ik}.\label{eq:DefConfigurationProjectorsEntries}
\end{equation}
 It follows immediately that the configuration projectors satisfy
the conditions of \emph{mutual exclusivity}, 
\begin{equation}
\projector_{i}\projector_{j}=\delta_{ij}\projector_{i},\label{eq:ConfigurationProjectorsMutuallyExclusive}
\end{equation}
 and \emph{completeness}, 
\begin{equation}
\sum^{N}_{i=1}\projector_{i}=\idmatrix,\label{eq:ConfigurationProjectorsComplete}
\end{equation}
 where again $\idmatrix$ is the $N\times N$ identity matrix. These
two conditions are the defining features of a projection-valued measure
(PVM) (Mackey 1952, 1957)\nocite{Mackey:1952irolcgi,Mackey:1957qmahs},
so the configuration projectors $\projector_{1},\dots,\projector_{N}$
naturally constitute a PVM.

Letting $\tr{\left(\cdots\right)}$ denote the usual trace, one can
then recast the identity $\stochasticmatrix_{ij}{\left(t\from0\right)}=\verts{\dynop_{ij}{\left(t\from0\right)}}^{2}$
from \eqref{eq:StochasticMatrixEntryFromAbsValSquare} as 
\begin{equation}
\keyeq{\stochasticmatrix_{ij}{\left(t\from0\right)}=\tr\parens{\dynop^{\adj}{\left(t\from0\right)}\projector_{i}\dynop{\left(t\from0\right)}\projector_{j}}.}\label{eq:DefDictionary}
\end{equation}
 This formula provides the \emph{dictionary} that translates between
indivisible stochastic processes, as represented by the left-hand
side, and the formalism of quantum-theoretic Hilbert spaces, as represented
by the right-hand side. This dictionary is the basic ingredient of
the \emph{stochastic\textendash quantum correspondence}.

\subsection{The Hilbert-Space Representation\label{subsec:The-Hilbert-Space-Representation}}

Given an initial standalone probability distribution, 
\begin{equation}
p_{j}{\left(0\right)}\defeq p{\left(j,0\right)},\label{eq:IndivisibleInitialStandaloneProbabilityDistribution}
\end{equation}
 perhaps capturing epistemic (\textquoteleft subjective credence\textquoteright )
uncertainty about the system's configuration $j$ at the initial time
$0$, one can define an initial \emph{density matrix} 
\begin{equation}
\densitymatrix{\left(0\right)}\defeq\diag{\dots,p_{j}{\left(0\right)},\dots}=\sum^{N}_{j=1}p_{j}{\left(0\right)}\projector_{j},\label{eq:DefInitialDensityMatrix}
\end{equation}
 which is a diagonal matrix whose diagonal entries are just the initial
probabilities $p_{j}{\left(0\right)}$, where $\projector_{j}$ are
the configuration projectors defined in \eqref{eq:DefConfigurationProjectorsAsDiagonalMatrices}.
One can then define a time-dependent density matrix by a congruence
transformation 
\begin{equation}
\densitymatrix{\left(t\right)}\defeq\dynop{\left(t\from0\right)}\densitymatrix{\left(0\right)}\dynop^{\adj}{\left(t\from0\right)},\label{eq:TimeDependentDensityMatrix}
\end{equation}
 which is not generically diagonal, but is self-adjoint, positive-semidefinite,
and has trace equal to $1$, due to the summation condition \eqref{eq:SumTimeEvOpAbsValSqEq1}
satisfied by the potential matrix $\dynop{\left(t\from0\right)}$.

If the system's density matrix $\densitymatrix{\left(t\right)}$ happens
to be rank-one, then it can be factorized as the outer product of
an $N\times1$ column vector $\Psi{\left(t\right)}$ with its complex-conjugate-transpose,
or adjoint, $\Psi^{\adj}{\left(t\right)}$: 
\begin{equation}
\densitymatrix{\left(t\right)}=\Psi{\left(t\right)}\Psi^{\adj}{\left(t\right)}\quad{\left[\textrm{if rank-one}\right]}.\label{eq:RankOneDensityMatrixFactorizedStateVector}
\end{equation}
 The \emph{state vector} or \emph{wave function} $\Psi{\left(t\right)}$
then evolves in time according to 
\begin{equation}
\Psi{\left(t\right)}=\dynop{\left(t\right)}\Psi{\left(0\right)},\label{eq:StateVectorTimeEv}
\end{equation}
 and is guaranteed to have unit-norm, in the sense that 
\begin{equation}
\sqrt{\Psi^{\adj}{\left(t\right)}\Psi{\left(t\right)}}=1.\label{eq:StateVectorUnitNorm}
\end{equation}
 However, $\Psi{\left(t\right)}$ is not unique, as it implicitly
depends on the choice of potential matrix $\dynop{\left(t\from0\right)}$,
and can therefore be altered by Schur\textendash Hadamard gauge transformations
\eqref{eq:DefLocalInTimeSchurHadamardGaugeTransformation}. Later
on, this paper will review a different kind of gauge invariance under
which $\Psi{\left(t\right)}$ is not unique. Even fixing both these
kinds of gauge invariance, the state vector $\Psi{\left(t\right)}$
is still only defined up to an overall phase factor. This paper's
demotion of state vectors from ontology to mathematical appurtenance
therefore has an additional benefit: this move naturally accounts
for the well-known fact that such overall phase factors are unphysical.\footnote{The author thanks an anonymous reviewer for suggesting an emphasis
of this point.}

In terms of the system's density matrix $\densitymatrix{\left(t\right)}$,
as defined in \eqref{eq:TimeDependentDensityMatrix}, one can express
the law of total probability \eqref{eq:IndivisibleStochasticProcessChapmanKolmogorovEquationsLawOfTotalProbability}
as 
\begin{equation}
p_{i}{\left(t\right)}=\tr\parens{\projector_{i}\densitymatrix{\left(t\right)}},\label{eq:FinalStandaloneProbabilitiesFromTraceDensityMatrix}
\end{equation}
 which is just the \emph{Born rule} for the configuration basis,
as expressed in terms of the system's density matrix $\densitymatrix{\left(t\right)}$
and the configuration projector $\projector_{i}$, as defined in \eqref{eq:DefConfigurationProjectorsAsDiagonalMatrices}.
The fact that this formula works out correctly, given the definition
\eqref{eq:TimeDependentDensityMatrix} of the system's density matrix,
is ultimately due to the linearity of the law of total probability
\eqref{eq:IndivisibleStochasticProcessChapmanKolmogorovEquationsLawOfTotalProbability}.
For the case of a rank-one density matrix, for which a state vector
$\Psi{\left(t\right)}$ is available, the Born rule \eqref{eq:FinalStandaloneProbabilitiesFromTraceDensityMatrix}
takes a form that is more commonly found in elementary textbooks,
\begin{equation}
p_{i}{\left(t\right)}=\verts{\Psi_{i}{\left(t\right)}}^{2},\label{eq:ConfigurationBornRuleFromStateVector}
\end{equation}
 where $\Psi_{i}{\left(t\right)}$ is the complex-valued $i$th entry
of $\Psi{\left(t\right)}$.

For any random variable $A{\left(t\right)}$ with possible values
$a{\left(i,t\right)}$, one can introduce a diagonal, self-adjoint
matrix, which, by a re-appropriation of notation, will be denoted
by the same symbol $A{\left(t\right)}$: 
\begin{equation}
A{\left(t\right)}\defeq\diag{\dots,a{\left(i,t\right)},\dots}=\sum^{N}_{i=1}a{\left(i,t\right)}\projector_{i}.\label{eq:DefBeableMatrix}
\end{equation}
 From straightforward linear algebra, one can rewrite the expectation
value \eqref{eq:DefExpectationValueRandomVariable} as 
\begin{equation}
\expectval{A{\left(t\right)}}=\tr{\left(A{\left(t\right)}\densitymatrix{\left(t\right)}\right)},\label{eq:ExpectationValueRandomVariableFromTrace}
\end{equation}
 or, for the rank-one case, as 
\begin{equation}
\expectval{A{\left(t\right)}}=\Psi^{\adj}{\left(t\right)}A{\left(t\right)}\Psi{\left(t\right)}.\label{eq:ExpectationValueRandomVariableFromInnerProduct}
\end{equation}

In keeping with Bell's terminology, random variables will now be called
\emph{be-ables}, or \emph{beables} (Bell 1973, 1976)\nocite{Bell:1973sao,Bell:1976ttolb}.
When the system is in any given configuration $i$, each beable $A{\left(t\right)}$
has a definite, underlying value $a{\left(i,t\right)}$. Bell intended
to distinguish beables from mere observables, which, by the Kochen\textendash Specker
theorem (Bell 1966; Kochen, Specker 1967)\nocite{Bell:1966otpohviqm,KochenSpecker:1967phvqm},
cannot all be beables. 

Observables that are not beables are represented by non-diagonal self-adjoint
operators that correspond to emergent patterns that show up when systems
interact with measuring devices, and will be called \emph{emergeables}
(Barandes 2025)\nocite{Barandes:2025tsqc}. The idea of emergeables
goes back to Bohr (Bohr 1935, Bell 1971)\nocite{Bohr:1935cqmdoprbcc,Bell:1971itthvq_2},
and also figures prominently in Bohmian mechanics (Bohm 1952b, Bell
1982, Daumer et al. 1996)\nocite{Bohm:1952siqtthvii,Bell:1982otipw,DaumerDurrGoldsteinZanghi:1996nrao}.
A system's beables and its emergeables collectively comprise the system's
overall non-commutative algebra of observables. When a measuring device
is properly modeled as one additional part of a larger stochastic
process, along the lines presented in the present paper, one can show
that at the end of the measurement process, the measuring device will
end up in one of its possible measurement-outcome configurations with
a stochastic probability that coincides with the standard Born rule,
whether the measuring device has been tuned to measure a beable or
an emergeable (Barandes 2025)\nocite{Barandes:2025tsqc}.

The Hilbert-space formulas introduced so far are all expressed in
what would conventionally be called the \emph{Schr{\"o}dinger picture}.
It can also be useful to introduce the \emph{Heisenberg picture},
which is defined according to 
\begin{equation}
\eqsbrace{\begin{aligned} & \densitymatrix^{H}\defeq\densitymatrix{\left(0\right)},\quad\Psi^{H}\defeq\Psi{\left(0\right)},\\
 & \quad A^{H}{\left(t\right)}\defeq\dynop^{\adj}{\left(t\from0\right)}A{\left(t\right)}\dynop{\left(t\from0\right)},
\end{aligned}
}\label{eq:DefHeisenbergPicture}
\end{equation}
 where $A^{H}{\left(t\right)}$, which is no longer generically diagonal,
now includes both a possible explicit dependence on time through its
magnitudes $a{\left(i,t\right)}$ as well as an implicit dependence
on time through the time-evolution operator $\dynop{\left(t\from0\right)}$.
In the Heisenberg picture, the probability formula \eqref{eq:FinalStandaloneProbabilitiesFromTraceDensityMatrix}
becomes\footnote{Note that for a generic time-evolution operator $\dynop{\left(t\from0\right)}$
that might not be unitary, the Heisenberg-picture version $\projector^{H}_{i}{\left(t\right)}\defeq\dynop^{\adj}{\left(t\from0\right)}\projector_{i}\dynop{\left(t\from0\right)}$
of a projector $\projector_{i}$ will not necessarily still be a projector.} 
\begin{equation}
p_{i}{\left(t\right)}=\tr\parens{\projector^{H}{\left(t\right)}\densitymatrix^{H}},\label{eq:FinalStandaloneProbabilitiesFromTraceDensityMatrixHeisenbergPicture}
\end{equation}
 and the formula \eqref{eq:ExpectationValueRandomVariableFromTrace}
for expectation values becomes 
\begin{equation}
\expectval{A{\left(t\right)}}=\tr\parens{A^{H}{\left(t\right)}\densitymatrix^{H}}.\label{eq:ExpectationValueRandomVariableFromTraceHeisenbergPicture}
\end{equation}

With the Heisenberg picture available, one can construct a general
example of an emergeable. Let $A$ be a time-independent beable, meaning
a random variable represented by a constant diagonal matrix \eqref{eq:DefBeableMatrix},
and consider the time derivative of its Heisenberg-picture counterpart
\begin{equation}
A^{H}{\left(t\right)}\defeq\dynop^{\adj}{\left(t\from0\right)}A{\left(t\right)}\dynop{\left(t\from0\right)},\label{eq:DefHeisenbergPictureEmergeable}
\end{equation}
 as defined for the time-evolution operator $\dynop{\left(t\from0\right)}$,
in the limit $t\to0$: 
\begin{equation}
\dot{A}\defeq\lim_{t\to0}\frac{dA^{H}{\left(t\right)}}{dt}=\dot{A}^{\adj}.\label{eq:DefTimeDerivEmergeableAtTimeZero}
\end{equation}
 This self-adjoint $N\times N$ matrix will generically be non-diagonal
and will therefore not generally commute with the original random
variable $A$ itself: 
\begin{equation}
\bracks{A,\dot{A}}\ne0.\label{eq:TimeDerivEmergeableNotCommuteWithRandomVar}
\end{equation}
 Nonetheless, the matrix $\dot{A}$ has physical uses. For example,
one has 
\begin{equation}
\tr\parens{\dot{A}\densitymatrix{\left(0\right)}}=\lim_{t\to0}\frac{d\expectval{A{\left(t\right)}}}{dt},\label{eq:TimeDerivExpectationValueFromTimeDerivEmergeable}
\end{equation}
 which is a perfectly meaningful physical quantity, even though the
time derivative of an expectation value is not necessarily the expectation
value of something physical.

The matrix $\dot{A}$ therefore resembles a random variable in some
ways, but incorporates stochastic dynamics directly into its definition
through the time-evolution operator $\dynop{\left(t\from0\right)}$,
and does not have a transparent interpretation at the level of the
underlying configuration space $\configspace$ of the given indivisible
stochastic process. Instead, $\dot{A}$ is an emergent amalgam of
kinematical and dynamical ingredients\textemdash that is, it is an
emergeable.

As a concrete example, consider a particle whose underlying \emph{position}
is regarded as a physical configuration, corresponding to some random
variable $A$. If the particle's dynamics is stochastic, in the sense
that the particle can be described in terms of an indivisible stochastic
process, then the particle's \emph{velocity} (or, equivalently, the
particle's \emph{momentum}) will not generally have a well-defined
value at all times, and will naturally be representable as an emergeable
$\dot{A}$ along the lines described here.

\subsection{Kraus Decompositions and Unitary Time Evolution\label{subsec:Kraus-Decompositions-and-Unitary-Time-Evolution}}

Given an $N\times N$ time-evolution operator $\dynop{\left(t\from0\right)}$
for the system, one can define a set of $N$ \emph{Kraus operators}
$\krausmatrix_{\beta}{\left(t\from0\right)}$, for $\beta=1,\dots,N$,
according to 
\begin{equation}
\krausmatrix_{\beta}{\left(t\from0\right)}\defeq\dynop{\left(t\from0\right)}\projector_{\beta}.\label{eq:DefKrausOperators}
\end{equation}
 Due to the summation condition $\sum^{N}_{i=1}\verts{\dynop_{ij}{\left(t\from0\right)}}^{2}=1$
from \eqref{eq:SumTimeEvOpAbsValSqEq1}, these operators satisfy the
Kraus identity 
\begin{equation}
\sum^{N}_{\beta=1}\krausmatrix^{\adj}_{\beta}{\left(t\from0\right)}\krausmatrix_{\beta}{\left(t\from0\right)}=\idmatrix,\label{eq:KrausIdentity}
\end{equation}
 as befits their name (Kraus 1971)\nocite{Kraus:1971gscqt}. In terms
of these Kraus operators, one can write the stochastic\textendash quantum
dictionary \eqref{eq:DefDictionary} as the Kraus decomposition 
\begin{equation}
\stochasticmatrix_{ij}{\left(t\from0\right)}=\sum^{N}_{\beta=1}\tr\parens{\krausmatrix^{\adj}_{\beta}{\left(t\from0\right)}\projector_{i}\krausmatrix_{\beta}{\left(t\from0\right)}\projector_{j}}.\label{eq:StochasticMatrixFromKrausDecomposition}
\end{equation}
 Consequently, one can write the time-evolution formula \eqref{eq:TimeDependentDensityMatrix}
for the system's density matrix $\densitymatrix{\left(t\right)}$
as the Kraus decomposition 
\begin{equation}
\densitymatrix{\left(t\right)}\defeq\sum^{N}_{\beta=1}\krausmatrix_{\beta}{\left(t\from0\right)}\densitymatrix{\left(0\right)}\krausmatrix^{\adj}_{\beta}{\left(t\from0\right)}.\label{eq:TimeDependentDensityMatrixFromKrausDecomposition}
\end{equation}

Like all the other mathematical objects in the Hilbert-space formulation,
the Kraus operators $\krausmatrix_{1}{\left(t\from0\right)}$, $\dots$,
$\krausmatrix_{N}{\left(t\from0\right)}$ are not unique. Notice also
that any number of $N\times N$ matrices satisfying the Kraus identity
\eqref{eq:KrausIdentity} and giving a suitable Kraus decomposition
akin to \eqref{eq:StochasticMatrixFromKrausDecomposition} are guaranteed
to yield a valid transition matrix $\stochasticmatrix{\left(t\from0\right)}$.
Said in another way, the preceding argument establishes the existence
but not the uniqueness of Kraus operators for any given time-evolution
operator $\dynop{\left(t\from0\right)}$.

As shown in other work (Barandes 2025, 2023)\nocite{Barandes:2025tsqc,Barandes:2023tsqt},
and as will be explained in detail later in the present paper, the
existence of these Kraus decompositions has an important implication.
Specifically, after an application, if necessary, of the Stinespring
dilation theorem (Stinespring 1955, Keyl 2002)\nocite{Stinespring:1955pfoc,Keyl:2002foqit},
which involves expanding the original $N$-dimensional Hilbert space
to a \textquoteleft dilated\textquoteright{} Hilbert space of dimension
no greater than $N^{3}$, one can always assume that the system's
time-evolution operator $\dynop{\left(t\from0\right)}$ is unitary,
meaning that 
\begin{equation}
\dynop{\left(t\from0\right)}=\timeevop{\left(t\from0\right)},\label{eq:UnitaryTimeEvOp}
\end{equation}
 with 
\begin{equation}
\timeevop^{\adj}{\left(t\from0\right)}=\timeevop^{-1}{\left(t\from0\right)}.\label{eq:TimeEvOpUnitarityCondition}
\end{equation}
 The preceding arguments therefore provide a first-principles motivation
for \emph{unitary time evolution}, or \emph{unitarity}, in quantum
theory.\footnote{It is widely known that if one is given a completely positive trace-preserving
(CPTP) map, or quantum channel, on a Hilbert space, then one can convert
that CPTP map into a unitary map via a Stinespring dilation. This
technique is sometimes called \textquoteleft going to the Church of
the Larger Hilbert space,\textquoteright{} a phrasing attributed to
John Smolin (Gottesman, Lo 2000)\nocite{GottesmanLo:2000fqctqs}.
The novel feature of the present paper's approach is that the starting
place was not a CPTP map on an already-given Hilbert space, but an
indivisible stochastic process that was given prior to the introduction
of a Hilbert space. The author thanks an anonymous reviewer for highlighting
this point.}

It follows that the basic relationship \eqref{eq:StochasticMatrixEntryFromAbsValSquare}
now takes the form 
\begin{equation}
\stochasticmatrix_{ij}{\left(t\from0\right)}=\verts{\timeevop_{ij}{\left(t\from0\right)}}^{2},\label{eq:StochasticMatrixEntryFromAbsValSquareUnitary}
\end{equation}
 and the dictionary \eqref{eq:DefDictionary} of the stochastic\textendash quantum
correspondence becomes 
\begin{equation}
\stochasticmatrix_{ij}{\left(t\from0\right)}=\tr\parens{\timeevop^{\adj}{\left(t\from0\right)}\projector_{i}\timeevop{\left(t\from0\right)}\projector_{j}}.\label{eq:DictionaryFromUnitary}
\end{equation}
 The time-evolution rule \eqref{eq:TimeDependentDensityMatrix} for
the system's density matrix is now 
\begin{equation}
\densitymatrix{\left(t\right)}\defeq\timeevop{\left(t\from0\right)}\densitymatrix{\left(0\right)}\timeevop^{\adj}{\left(t\from0\right)},\label{eq:TimeDependentDensityMatrixUnitary}
\end{equation}
 and if the density matrix is rank-one, so that it factorizes as $\densitymatrix{\left(t\right)}=\Psi{\left(t\right)}\Psi^{\adj}{\left(t\right)}$
as in \eqref{eq:RankOneDensityMatrixFactorizedStateVector} for a
state vector $\Psi{\left(t\right)}$, then the state vector's time-evolution
rule \eqref{eq:StateVectorTimeEv} is 
\begin{equation}
\Psi{\left(t\right)}=\timeevop{\left(t\from0\right)}\Psi{\left(0\right)}.\label{eq:StateVectorTimeEvUnitary}
\end{equation}
 The Heisenberg picture \eqref{eq:DefHeisenbergPicture} is then 
\begin{equation}
\eqsbrace{\begin{aligned} & \densitymatrix^{H}\defeq\densitymatrix{\left(0\right)},\quad\Psi^{H}\defeq\Psi{\left(0\right)},\\
 & \quad A^{H}{\left(t\right)}\defeq\timeevop^{\adj}{\left(t\from0\right)}A{\left(t\right)}\timeevop{\left(t\from0\right)}.
\end{aligned}
}\label{eq:DefHeisenbergPictureUnitary}
\end{equation}

In general, an $N\times N$ matrix is called \emph{unistochastic}
if its individual entries are expressible as the modulus-squares of
the corresponding entries of an $N\times N$ unitary matrix. It follows
that \eqref{eq:StochasticMatrixEntryFromAbsValSquareUnitary} is just
the statement that the system's transition matrix $\stochasticmatrix{\left(t\from0\right)}$
can be taken to be unistochastic, and will therefore be said to describe
a \emph{unistochastic process}.

The preceding analysis implies that an indivisible stochastic process
can be viewed either as a unistochastic process itself, or (if a nontrivial
dilation was required) as a \emph{subsystem} of a unistochastic process.
This statement is called the \emph{stochastic\textendash quantum theorem}
(Barandes 2023)\nocite{Barandes:2023tsqt}.

Note that a unitary time-evolution operator $\timeevop{\left(t\from0\right)}$
will not generically remain unitary under arbitrary Schur\textendash Hadamard
gauge transformations \eqref{eq:DefSchurHadamardGaugeTransformationEntries}.
Hence, writing a unistochastic transition matrix $\stochasticmatrix{\left(t\from0\right)}$
in terms of a unitary time-evolution operator $\timeevop{\left(t\from0\right)}$
corresponds to making a gauge choice\textemdash or, somewhat more
precisely, to a partial fixing of the gauge freedom \eqref{eq:DefSchurHadamardGaugeTransformationEntries}.

If the unitary time-evolution operator $\timeevop{\left(t\from0\right)}$
is an appropriately differentiable function of the target time $t$,
then one can define a corresponding self-adjoint matrix $H{\left(t\right)}$,
called the \emph{Hamiltonian}, according to 
\begin{equation}
H{\left(t\right)}\defeq i\hbar\frac{\partial\timeevop{\left(t\from0\right)}}{\partial t}\timeevop^{\adj}{\left(t\from0\right)}=H^{\adj}{\left(t\right)}.\label{eq:DefHamiltonian}
\end{equation}
  Here the factor of $i$ ensures self-adjointness. The reduced Planck
constant $\hbar$ is introduced to give $H{\left(t\right)}$ measurement
units of energy, with an explicit value fixed by the definition of
one's system of measurement units. It follows immediately that the
system's density matrix satisfies the \emph{von Neumann equation},
\begin{equation}
i\hbar\frac{\partial\densitymatrix{\left(t\right)}}{\partial t}={\left[H{\left(t\right)},\densitymatrix{\left(t\right)}\right]},\label{eq:VonNeumannEq}
\end{equation}
 the system's state vector (which exists if the density matrix is
rank-one) satisfies the \emph{Schr{\"o}dinger equation}, 
\begin{equation}
i\hbar\frac{\partial\Psi{\left(t\right)}}{\partial t}=H{\left(t\right)}\Psi{\left(t\right)},\label{eq:SchrodingerEq}
\end{equation}
 expectation values satisfy the \emph{Ehrenfest equation}, 
\begin{equation}
\frac{d\expectval{A{\left(t\right)}}}{dt}=\expectval{\frac{i}{\hbar}{\left[H{\left(t\right)},A{\left(t\right)}\right]}}+\expectval{\frac{\partial A{\left(t\right)}}{\partial t}},\label{eq:EhrenfestEq}
\end{equation}
 and Heisenberg-picture random variables $A^{H}{\left(t\right)}$
evolve according to the \emph{Heisenberg equation of motion}, 
\begin{equation}
\frac{dA^{H}{\left(t\right)}}{dt}=\frac{i}{\hbar}\bracks{H^{H}{\left(t\right)},A^{H}{\left(t\right)}}+{\left(\frac{\partial A{\left(t\right)}}{\partial t}\right)}^{H},\label{eq:HeisenbergEquationOfMotion}
\end{equation}
 where the matrix $H^{H}{\left(t\right)}$ appearing in the Heisenberg
equation of motion \eqref{eq:HeisenbergEquationOfMotion} is the Hamiltonian
expressed in the Heisenberg picture. Note that the brackets here are
\emph{commutators}, ${\left[A,B\right]}\defeq AB-BA$, not Poisson
brackets, and that these equations are not mere analogies rooted in
a classical Liouville picture.

This paper has reviewed how the stochastic\textendash quantum correspondence
allows one to write an indivisible stochastic process in a Hilbert-space
form describing an associated quantum system. From this standpoint,
one begins with a choice of configuration space appropriate to the
ontology of the kind of stochastic process that one is trying to model,
and then one builds a Hilbert-space representation in which that configuration
space corresponds to a preferred basis\textemdash the configuration
basis. Meanwhile, the nomological transition probabilities of the
stochastic process end up represented in terms of a time-evolution
operator on the Hilbert space. Then various features of the underlying
stochastic process show up in the Hilbert-space formulation in various
exotic-looking forms, such as in the case of indivisibility manifesting
on the Hilbert-space side as interference effects (Barandes 2025)\nocite{Barandes:2025tsqc}.

By reading the dictionary \eqref{eq:DictionaryFromUnitary} in the
other logical direction, a quantum system evolving unitarily can be
understood as an indivisible stochastic process in disguise. The correspondence
between indivisible stochastic processes and Hilbert-space quantum
theory is akin to the correspondence between classical Newtonian systems
and Hamiltonian mechanics. In each pairing, the former (respectively,
indivisible stochastic processes and Newtonian systems) give a clearer
picture of the ontology, whereas the latter (respectively, Hilbert-space
quantum theory and Hamiltonian mechanics) provide powerful mathematical
tools for constructing new dynamics and calculating predictions. In
both cases, the correspondence is many-to-one in both directions.

That last sentence is particularly important. Due to all the many
choices involved in constructing a Hilbert space for a given indivisible
stochastic process, a given indivisible stochastic process can have
infinitely many different Hilbert-space representations. Going the
other way, a given Hilbert-space quantum system will represent infinitely
many qualitatively different indivisible stochastic processes, where
each different orthonormal basis for the Hilbert space could end up
defining a configuration space for a different indivisible stochastic
process.

This situation is much like the relationship between Newtonian mechanics
and the classical Hamiltonian formulation, as emphasized, for instance
by Curiel (2014)\nocite{Curiel:2014cmiliinh}. If one is given just
a classical Hamiltonian formulation in the abstract, but not any information
specifying which set of canonical frames ${\left(q_{1},\dots,q_{n},p_{1},\dots,p_{n}\right)}$
are physical among all canonical frames connected by canonical transformations,
then there will not, in general, be a unique such choice of \textquoteleft ontological\textquoteright{}
canonical frame. Even for the simple example of a classical harmonic
oscillator, there is no way to know whether $q$ is position and $p$
is momentum, with mass $m$ and spring constant $k$, or whether $p$
is position and $-q$ is momentum, with mass $1/k$ and spring constant
$1/m$. Indeed, this flexibility in the connection between formalism
and ontology is one of the powerful things about the classical Hamiltonian
formulation, and means that calculations done in the classical Hamiltonian
formulation can be applied productively to many qualitatively different
kinds of Newtonian systems. All these considerations are ultimately
a manifestation of the famous underdetermination problem in the philosophy
of science.

All that said, when starting from an abstract Hilbert-space formulation,
some basic guidelines for picking a configuration space are still
presumably applicable. One can favor a choice in which the degrees
of freedom for the configuration space are local in physical space,
decompose in a mereologically sensible way among subsystems, lead
to the quantum Hamiltonian having an appropriately local-looking form,
exhibit suitably rapid decoherence to their associated configuration
basis, and are sufficiently rich that all the readings that appear
on measuring devices (or in the brains of external observers) can
supervene on them. As with classical physics, one can also try to
pick degrees of freedom that lead to an intuitive and qualitatively
perspicuous connection with the system's observables\textemdash for
example, if a system's observables show up as pointlike dots on a
detection screen, then this pattern of observations might suggest
picking particle arrangements as the configurations. In the end, however,
the stochastic\textendash quantum correspondence does not provide
any miraculous new ways to gain epistemic access to a system's true
underlying ontology, so scientific underdetermination lives another
day.

\subsection{Foldy\textendash Wouthuysen Gauge Transformations\label{subsec:Foldy-Wouthuysen-Gauge-Transformations}}

The Hilbert-space formulation has \emph{another} form of gauge invariance,
which appears to have first been written down by Foldy and Wouthuysen
in a 1950 paper on the Dirac equation (Foldy, Wouthuysen)\nocite{FoldyWouthuysen:1950otdtos12painrl}.\footnote{Foldy and Wouthuysen originally used the term \emph{canonical transformation}
rather than \emph{gauge transformation}, and wrote the operator $V$
instead as $\exp{\left(iS\right)}$, where $S=S{\left(t\right)}$
was a time-dependent, Hermitian operator. These transformations were
later described in the textbook on quantum mechanics by Messiah (1958,
Volume 1, Chapter XX, Section 33)\nocite{Messiah:1961qm}, and also
in the textbook on relativistic quantum mechanics by Bjorken and Drell
(1964, Chapter 4)\nocite{BjorkenDrell:1964rqm}. They were a particular
focus of a paper by Goldman (1977)\nocite{Goldman:1977gitdfwtatph},
and were apparently discovered independently by Brown (1999)\nocite{Brown:1999aooiqm},
inspired by his study of transformations of the Schr{\"o}dinger equation
between inertial and non-inertial reference frames.} Working with a generic time-evolution operator $\dynop{\left(t\from0\right)}$
that may or may not be unitary, and letting $V{\left(t\right)}$ be
a time-dependent unitary matrix, the following transformation is a
gauge invariance of the entire Hilbert-space formulation, leaving
all probabilities $p_{i}{\left(t\right)}$, expectation values $\expectval{A{\left(t\right)}}$,
and the transition matrix $\stochasticmatrix{\left(t\from0\right)}$
as a whole unchanged:\footnote{One should be mindful of the appearance of the initial time $0$ in
$V^{\adj}{\left(0\right)}$ in the transformation rule for $\dynop{\left(t\from0\right)}$.} 
\begin{equation}
\eqsbrace{\begin{aligned}\densitymatrix{\left(t\right)} & \mapsto\densitymatrix_{V}{\left(t\right)}\defeq V{\left(t\right)}\densitymatrix{\left(t\right)}V^{\adj}{\left(t\right)},\\
\Psi{\left(t\right)} & \mapsto\Psi_{V}{\left(t\right)}\defeq V{\left(t\right)}\Psi{\left(t\right)},\\
A{\left(t\right)} & \mapsto A_{V}{\left(t\right)}\defeq V{\left(t\right)}A{\left(t\right)}V^{\adj}{\left(t\right)},\\
\dynop{\left(t\from0\right)} & \mapsto\dynop_{V}{\left(t\from0\right)}\defeq V{\left(t\right)}\dynop{\left(t\from0\right)}V^{\adj}{\left(0\right)}.
\end{aligned}
}\label{eq:DefLocalInTimeUnitaryTransformation}
\end{equation}

If the unitary matrix $V{\left(t\right)}$ is taken to be time-\emph{independent},
then the Foldy\textendash Wouthuysen gauge transformation \eqref{eq:DefLocalInTimeUnitaryTransformation}
is merely a simple change of orthonormal basis. However, if $V{\left(t\right)}$
depends nontrivially on time, and if one regards the system's Hilbert
space at each moment in time as a fiber over a one-dimensional base
manifold parameterized by the time coordinate $t$, then $V{\left(t\right)}$
represents a local-in-time, unitary transformation of each individual
Hilbert-space fiber. In particular, any given time-dependent state
vector $\Psi{\left(t\right)}$, regarded as a trajectory through the
system's Hilbert space, can be mapped to any other trajectory by a
suitable choice of time-dependent unitary matrix $V{\left(t\right)}$,
so trajectories in the Hilbert space do not describe gauge-invariant
facts.

These formulas make manifest that the Hilbert-space formulation of
an indivisible stochastic process is ultimately a collection of gauge-dependent
quantities, or gauge variables. In any physical theory, one does not
typically try to assign gauge variables an ontological meaning. The
familiar ingredients of the Hilbert-space formulation are highly gauge-dependent
according to both kinds of gauge transformations described in the
present paper\textemdash Schur\textendash Hadamard gauge transformations
\eqref{eq:DefLocalInTimeSchurHadamardGaugeTransformation} and Foldy\textendash Wouthuysen
gauge transformations \eqref{eq:DefLocalInTimeUnitaryTransformation}.
Hence, although a Hilbert-space formulation may be extremely useful
for constructing stochastic dynamics or for carrying out calculations,
one might rightly be skeptical about trying to assign direct physical
meanings to its mathematical ingredients, or suspicious of any interpretation
of quantum theory based on reifying, say, wave functions or density
matrices as parts of the ontological furniture of reality. 

Intriguingly, if the system's time-evolution operator $\dynop{\left(t\from0\right)}=\timeevop{\left(t\from0\right)}$
can be taken to be unitary, then under Foldy\textendash Wouthuysen
gauge transformation defined by \eqref{eq:DefLocalInTimeUnitaryTransformation},
the Hamiltonian $H{\left(t\right)}$ defined in \eqref{eq:DefHamiltonian}
transforms precisely as a non-Abelian gauge potential:\footnote{For pedagogical treatments of non-Abelian gauge theories, see Peskin,
Schroeder (1999) or Weinberg (1996)\nocite{PeskinSchroeder:1995iqft,Weinberg:1996tqtfii}.} 
\begin{equation}
\eqsbrace{\begin{aligned} & H{\left(t\right)}\mapsto H_{V}{\left(t\right)}\\
 & \qquad=V{\left(t\right)}H{\left(t\right)}V^{\adj}{\left(t\right)}-i\hbar V{\left(t\right)}\frac{\partial V^{\adj}{\left(t\right)}}{\partial t}.
\end{aligned}
\negmedspace\negmedspace\negmedspace\negmedspace\negmedspace\negmedspace}\label{eq:HamiltonianNonAbelianTransformation}
\end{equation}
 This transformation behavior makes clear that a Hamiltonian is not
a gauge-invariant observable, even though it may happen to coincide
with various observables according to particular gauge choices. For
example, for a simple system of non-relativistic point particles,
there is generically a gauge choice in which the Hamiltonian is equal
to the sum of the observables representing the system's kinetic and
potential energies, but the Hamiltonian may look different according
to other gauge choices.

Observe that one can rewrite the Schr{\"o}dinger equation \eqref{eq:SchrodingerEq}
as 
\begin{equation}
\mathcal{D}{\left(t\right)}\Psi{\left(t\right)}=0.\label{eq:GaugeCovariantSchrodiginerEq}
\end{equation}
 Here $\mathcal{D}{\left(t\right)}$ is a gauge-covariant derivative
defined according to 
\begin{equation}
\mathcal{D}{\left(t\right)}\defeq\idmatrix\frac{\partial}{\partial t}+\frac{i}{\hbar}H{\left(t\right)},\label{eq:DefGaugeCovDeriv}
\end{equation}
 which maintains its form under Foldy\textendash Wouthuysen gauge
transformations \eqref{eq:DefLocalInTimeUnitaryTransformation}, in
the sense that 
\begin{equation}
V{\left(t\right)}{\left[\idmatrix\frac{\partial}{\partial t}+\frac{i}{\hbar}H{\left(t\right)}\right]}{\left(\cdots\right)}={\left[\idmatrix\frac{\partial}{\partial t}+\frac{i}{\hbar}H_{V}{\left(t\right)}\right]}{\left[V{\left(t\right)}{\left(\cdots\right)}\right]}.\label{eq:GaugeCovDerivTransf}
\end{equation}

Notice that if one picks the Foldy\textendash Wouthuysen gauge-transformation
matrix $V{\left(t\right)}$ to be the adjoint of the unistochastic
process's time-evolution operator $\timeevop{\left(t\from0\right)}$,
\begin{equation}
V{\left(t\right)}\defeq\timeevop^{\adj}{\left(t\from0\right)},\label{eq:DefLocalInTimeUnitaryTransformationForHeisenbergPicture}
\end{equation}
 then the Hamiltonian precisely vanishes: 
\begin{equation}
H_{V}{\left(t\right)}=0.\label{eq:DefLocalInTimeUnitaryTransformationForHeisenbergPictureHamiltonianZero}
\end{equation}
 This choice of gauge is nothing other than the definition \eqref{eq:DefHeisenbergPictureUnitary}
of the Heisenberg picture. Foldy\textendash Wouthuysen gauge transformations
\eqref{eq:DefLocalInTimeUnitaryTransformation} can therefore be viewed
as generalized changes of time-evolution picture.\footnote{The fact that one can set $H_{V}{\left(t\right)}=0$ for all $t$
is a manifestation of the fact that the fiber bundle in this case,
consisting of copies of the system's Hilbert space fibered over a
one-dimensional base manifold parameterized by the time $t$, has
vanishing curvature.}

\section{Further Implications\label{sec:Further-Implications}}

\subsection{Dynamical Symmetries\label{subsec:Dynamical-Symmetries}}

The stochastic\textendash quantum correspondence developed in this
paper provides new ways of thinking about \emph{dynamical symmetries}
in quantum theory, meaning transformations that leave the dynamics
invariant. Going in the other direction, the stochastic\textendash quantum
correspondence also makes it more straightforward to impose dynamical
symmetries systematically as constraints in the construction of a
given indivisible stochastic process.

In the textbook approach to quantum theory, based on the Dirac\textendash von
Neumann axioms, the only clearly ontological ingredients with contingent
physical configurations are external observers, measuring devices,
and their measurement results. Symmetry transformations are then most
naturally phrased in \emph{passive} terms, as changes in the perspective
of an external observer or measuring device.\footnote{Weinberg, for instance, writes that ``A symmetry transformation is
a change in our point of view that does not change the results of
possible experiments'' (Weinberg 1996, Section 2.2, pp. 50\textendash 51)\nocite{Weinberg:1996tqtfi}.} 

From this standpoint, it is not clear what it would mean to carry
out an \emph{active} symmetry transformation on the quantum system
itself. From the standpoint of indivisible quantum theory, by contrast,
active symmetry transformations have essentially the same meaning
as they do in classical physics\textemdash as changes to the given
system's physical states or configurations.

As an example, an invertible transformation of a system's configurations
$i=1,\dots,N$ could take the form of a permutation transformation
$\mapping{\sigma}{\set{1,\dots,N}}{\set{1,\dots,N}}$ of the configuration
projectors \eqref{eq:DefConfigurationProjectorsAsDiagonalMatrices}:
\begin{equation}
\eqsbrace{\begin{aligned} & \projector_{i}\mapsto\projector_{\sigma{\left(i\right)}},\\
 & \quad\textrm{with }\set{\sigma{\left(1\right)},\dots,\sigma{\left(N\right)}}=\set{1,\dots,N}.
\end{aligned}
}\label{eq:ClassicalSymmetryTransformationAsPermutation}
\end{equation}
 More generally, an invertible transformation that alters the system's
configurations in some more fundamental way should still allow for
a Hilbert-space description in which the system's new configurations
are again represented by a PVM. If $\projector_{i}$ represented the
$i$th configuration of the system before the transformation, then
let $\tilde{\projector}_{i}$ represent the corresponding configuration
after the transformation. An invertible transformation between two
PVMs $\projector_{1},\dots,\projector_{N}$ and $\tilde{\projector}_{1},\dots,\tilde{\projector}_{N}$
is always a similarity transformation of the form 
\begin{equation}
\projector_{i}\mapsto\tilde{\projector}_{i}\defeq V^{\adj}\projector_{i}V,\label{eq:DefGeneralSymmetryTransformation}
\end{equation}
 where $V$ is some unitary operator.\footnote{Proof: Let $e_{1},\dots,e_{N}$ be the orthonormal configuration basis
\eqref{eq:DefConfigurationBasis}, with $e^{\adj}_{i}e_{j}=\delta_{ij}$
and $e_{i}e^{\adj}_{i}=\projector_{i}$ as in \eqref{eq:ConfigurationBasisOrthonormalComplete},
and let $\tilde{e}_{1},\dots,\tilde{e}_{N}$ be an orthonormal basis
related to the new projectors $\tilde{\projector}_{i}$ in the analogous
way, with $\tilde{e}^{\adj}_{i}\tilde{e}_{j}=\delta_{ij}$ and $\tilde{e}_{i}\tilde{e}^{\adj}_{i}=\tilde{\projector}_{i}$.
Then the $N\times N$ matrix defined by $V\defeq\sum_{i}e_{i}\tilde{e}^{\adj}_{i}$
is unitary and satisfies $V^{\adj}\projector_{i}V=\tilde{\projector}_{i}$.
Going the other way, if $V$ is an $N\times N$ unitary matrix, then
the $N\times N$ matrices defined for $i=1,\dots,N$ by $\tilde{\projector}_{i}\defeq V^{\adj}\projector_{i}V$
are guaranteed to constitute a PVM.~QED} This similarity transformation reduces to the simple transformation
\eqref{eq:ClassicalSymmetryTransformationAsPermutation} if and only
if $V$ is a permutation matrix.

The next step is to determine what condition ensures that the more
general transformation \eqref{eq:DefGeneralSymmetryTransformation}
is a dynamical symmetry, meaning that it leaves the transition matrix\textemdash and
thus the stochastic dynamics\textemdash invariant. The required condition
is precisely that the right-hand side of the stochastic\textendash quantum
dictionary \eqref{eq:DefDictionary} should remain unchanged: 
\begin{equation}
\tr\parens{\dynop^{\adj}{\left(t\from0\right)}\tilde{\projector}_{i}\dynop{\left(t\from0\right)}\tilde{\projector}_{j}}=\tr\parens{\dynop^{\adj}{\left(t\from0\right)}\projector_{i}\dynop{\left(t\from0\right)}\projector_{j}}.\label{eq:DynamicalSymmetryDictionaryTransformedProjectors}
\end{equation}
 This condition is equivalent to the statement that 
\begin{equation}
\tr\parens{\tilde{\dynop}^{\adj}{\left(t\from0\right)}\projector_{i}\tilde{\dynop}{\left(t\from0\right)}\projector_{j}}=\tr\parens{\dynop^{\adj}{\left(t\from0\right)}\projector_{i}\dynop{\left(t\from0\right)}\projector_{j}},\label{eq:DynamicalSymmetryDictionaryTransformedTimeEvOp}
\end{equation}
 where 
\begin{equation}
\tilde{\dynop}{\left(t\from0\right)}\defeq V\dynop{\left(t\from0\right)}V^{\adj}.\label{eq:SymmetryTransformedTimeEvOp}
\end{equation}
 Re-expressing both sides of the equivalent condition \eqref{eq:DynamicalSymmetryDictionaryTransformedTimeEvOp}
in terms of modulus-squared values, as in \eqref{eq:StochasticMatrixEntryFromAbsValSquare},
one sees that \eqref{eq:SymmetryTransformedTimeEvOp} is a dynamical
symmetry precisely if 
\begin{equation}
\verts{\tilde{\dynop}_{ij}{\left(t\from0\right)}}^{2}=\verts{\dynop_{ij}{\left(t\from0\right)}}^{2}.\label{eq:DynamicalSymmetryTransfAbsValSqCondition}
\end{equation}

It follows immediately that $\tilde{\dynop}{\left(t\from0\right)}$
can differ from $\dynop{\left(t\from0\right)}$ by at most a Schur\textendash Hadamard
gauge transformation \eqref{eq:DefLocalInTimeSchurHadamardGaugeTransformation},
so a necessary and sufficient condition for a unitary matrix $V$
to give a dynamical symmetry is that 
\begin{equation}
V\dynop{\left(t\from0\right)}V^{\adj}=\dynop{\left(t\from0\right)}\hadamardprod\begin{pmatrix}e^{i\theta_{11}{\left(t\right)}} & e^{i\theta_{12}{\left(t\right)}}\\
e^{i\theta_{21}{\left(t\right)}} & \ddots\\
 &  & e^{i\theta_{NN}{\left(t\right)}}
\end{pmatrix}.\label{eq:DynamicalSymmetryAsLocalInTimeEntrywisePhaseTransformation}
\end{equation}
 As special cases, this condition includes \emph{unitary} dynamical
symmetries, 
\begin{equation}
V\dynop{\left(t\from0\right)}V^{\adj}=\dynop{\left(t\from0\right)},\label{eq:UnitaryDynamicalSymmetr}
\end{equation}
 as well as \emph{anti-unitary} dynamical symmetries, 
\begin{equation}
V\dynop{\left(t\from0\right)}V^{\adj}=\overconj{\dynop{\left(t\from0\right)}},\label{eq:AntiUnitaryDynamicalSymmetry}
\end{equation}
 where the overline notation denotes complex conjugation, which is
ultimately just a change of phases.

For the specific case of an anti-unitary dynamical symmetry, note
that if one redefines $V\mapsto\overconj V$, which is still unitary,
then one can re-express \eqref{eq:AntiUnitaryDynamicalSymmetry} in
the somewhat more conventional form 
\begin{equation}
VK\dynop{\left(t\from0\right)}KV^{\adj}=\dynop{\left(t\from0\right)}.\label{eq:AntiUnitaryDynamicalSymmetryUsingComplexConjOp}
\end{equation}
 Here $K$ denotes the complex-conjugation operator, meaning that
$K$ is an involution, 
\begin{equation}
K^{2}=1,\label{eq:ComplexConjugationOperatorInvolution}
\end{equation}
 and, for any $N\times N$ matrix $X$, one has 
\begin{equation}
KXK=\overconj X.\label{eq:ComplexConjugationOperatorOnMatrix}
\end{equation}
 The composite operator $VK$ as a whole is then said to be an \emph{anti-unitary operator}.
Anti-unitary operators play an important role in describing time-reversal
symmetries.\footnote{Intriguingly, because $K$ anticommutes with $i$, meaning that $Ki=-iK$,
the three mathematical objects $i$, $K$, and $iK$ satisfy $-i^{2}=K^{2}={\left(iK\right)}^{2}=iK{\left(iK\right)}=1$,
and therefore generate a Clifford algebra isomorphic to the \emph{pseudo-quaternions}
(Stueckelberg 1960)\nocite{Stueckelberg:1960qtirhs}. In a sense,
then, the Hilbert spaces of quantum systems are actually defined not
over the complex numbers alone, but over the pseudo-quaternions, although
$K$ is not typically involved in defining observables.}

If $\dynop{\left(t\from0\right)}=\timeevop{\left(t\from0\right)}$
is unitary, then $V\dynop{\left(t\from0\right)}V^{\adj}$ will likewise
be unitary. In that case, suppose either that $V$ is continuously
connected to the identity matrix $\idmatrix$ by some smooth parameter,
with a corresponding self-adjoint generator $G=G^{\adj}$, or, alternatively,
that $V$ is an involution, meaning that $V^{2}=\idmatrix$, in which
case $G\defeq V=V^{\adj}$ is \emph{itself} self-adjoint. Either way,
$G$ is self-adjoint, and therefore represents a candidate observable,
at least as an emergeable, so the expectation value $\expectval{G{\left(t\right)}}$
is an empirically meaningful quantity at the level of measurement
processes. If $G$ commutes with $\timeevop{\left(t\from0\right)}$,
then Noether's theorem easily follows as the statement that this expectation
value is constant in time, or conserved: 
\begin{equation}
\expectval{G{\left(t\right)}}=\tr\parens{G\timeevop{\left(t\from0\right)}\densitymatrix{\left(0\right)}\timeevop^{\adj}{\left(t\from0\right)}}=\expectval{G{\left(0\right)}}.\label{eq:NoethersTheorem}
\end{equation}

As a potentially new result, the condition \eqref{eq:DynamicalSymmetryAsLocalInTimeEntrywisePhaseTransformation}
may also open up the possibility of dynamical symmetries that are
distinct from the unitary and anti-unitary cases. This possibility
does not contradict Wigner's theorem (Wigner 1931)\nocite{Wigner:1931guiaadqmda},
because the discussion at this point is pitched at the level of an
\emph{entire system}, which may explicitly include measuring devices
as subsystems. As explained in other work (Barandes 2025)\nocite{Barandes:2025tsqc},
if one suitably models an entire system that includes a measuring
device as an overall unistochastic process, then the measuring device
will end up in one of its possible measurement-outcome configurations
with the appropriate Born-rule probability for whatever observable
is measured\textemdash whether a beable (represented by a self-adjoint
matrix that is diagonal in the configuration basis) or an emergeable
(represented by a non-diagonal self-adjoint matrix). Relative phases
in this overall unistochastic process are immaterial gauge variables.

However, for Wigner's theorem, as with the traditional textbook formulation
of the Dirac\textendash von Neumann axioms, one assumes that an \emph{implicit}
measuring device is left out of the Hilbert-space formalism, and one
can appeal to the Born rule and the collapse rule as bare posits.
In particular, the implicit measuring device does not participate
in the invertible transformation in question, and one needs to be
mindful of relative phases in the quantum system that the measuring
device is investigating. Because beables and emergeables in the quantum
system appear on an essentially similar footing as far as the implicit
measuring device is concerned, one therefore needs to specialize the
dynamical symmetries \eqref{eq:DynamicalSymmetryAsLocalInTimeEntrywisePhaseTransformation}
to a smaller class of transformations, which will be called \emph{Wigner symmetries}.

After carrying out a Wigner symmetry, any observables\textemdash beables
or emergeables\textemdash exhibited by the quantum system should still
be available to measuring devices. Each observable, being represented
by a self-adjoint operator, corresponds to a PVM of its own, so if
$\projector^{\prime}_{\alpha}$ is an element of this PVM before the
symmetry transformation, then let $\tilde{\projector}^{\prime}_{\alpha}$
denote the corresponding PVM element after carrying out the Wigner
symmetry. By precisely the same reasoning that led to the result \eqref{eq:DefGeneralSymmetryTransformation},
it follows that these two PVMs are related by some unitary operator
$V^{\prime}$: 
\begin{equation}
\tilde{\projector}^{\prime}_{\alpha}\mapsto\tilde{\projector}^{\prime}_{\alpha}\defeq V^{\prime\adj}\projector^{\prime}_{\alpha}V^{\prime}.\label{eq:DefGeneralSymmetryTransformationForOtherBasis}
\end{equation}
 Ensuring that the Born rule is invariant in the observable's orthonormal
basis leads to the condition 
\begin{equation}
\tr\parens{\dynop^{\adj}{\left(t\from0\right)}\tilde{\projector}^{\prime}_{\alpha}\dynop{\left(t\from0\right)}\tilde{\projector}^{\prime}_{\beta}}=\tr\parens{\dynop^{\adj}{\left(t\from0\right)}\projector^{\prime}_{\alpha}\dynop{\left(t\from0\right)}\projector^{\prime}_{\beta}},\label{eq:DynamicalSymmetryDictionaryTransformedProjectorsOtherBasis}
\end{equation}
 which ultimately leads to the statement that 
\begin{equation}
\verts{\tilde{\dynop}^{\prime}_{\alpha\beta}{\left(t\from0\right)}}^{2}=\verts{\dynop^{\prime}_{\alpha\beta}{\left(t\from0\right)}}^{2},\label{eq:DynamicalSymmetryTransfAbsValSqConditionOtherBasis}
\end{equation}
 a formula that must hold for all choices of orthonormal basis, and
not merely for the configuration basis as in \eqref{eq:DynamicalSymmetryTransfAbsValSqCondition}.
This stronger constraint is equivalent to the usual starting point
for Wigner's theorem,\footnote{See, for instance, eq. (2.A.1) in Weinberg's textbook on quantum field
theory (Weinberg 1996, Chapter 2, Appendix A)\nocite{Weinberg:1996tqtfi}.} whose conclusion is that Wigner symmetries must be implemented either
by unitary or anti-unitary operators. 

\subsection{Dilations\label{subsec:Dilations}}

In most textbook treatments of quantum theory, a quantum system is
axiomatically defined as a particular Hilbert space, together with
a preferred set of self-adjoint operators designated as observables
with predetermined physical meanings, along with a particular Hamiltonian
to define the system's time evolution.\footnote{In some circumstances, it may turn out to be more convenient to define
a quantum system by a formal C{*}-algebra of observables alone, without
picking a specific Hilbert-space representation (Haag 1993; Clifton,
Halvorson 2001; Strocchi 2008; Feintzeig 2016)\nocite{Haag:1993lqpfpa,CliftonHalvorson:2001arqripciqft,Strocchi:2008aittmsoqm,Feintzeig:2016oamiqt}.
Indeed, some have argued (see, for example, Ruetsche 2013)\nocite{Ruetsche:2013ueape}
that there are reasons beyond mere convenience to favor C{*}-algebras
over Hilbert-space representations, due, in part, to the well-known
problem that many C{*}-algebras have faithful but unitarily inequivalent
Hilbert-space representations that are empirically indistinguishable
at any finite level of experimental resolution, at least based on
Fell's theorem (Fell 1960)\nocite{Fell:1960tdpocsa}. The author thanks
an anonymous reviewer's suggestion to emphasize this point.} From that point of view, modifying a system's Hilbert-space formulation
in any nontrivial way would necessarily mean fundamentally modifying
the system itself.

From the alternative point of view developed in this paper, by contrast,
a Hilbert-space formulation is merely a collection of mathematical
tools for constructing the dynamics of a given indivisible stochastic
process or carrying out calculations more efficiently, and is no more
fundamental than a Lagrangian or Hamiltonian description of a Newtonian
system. The indivisible stochastic process itself is ultimately defined
by a configuration space and a dynamical law that stand apart from
any arbitrary choice of Hilbert-space formulation. As a consequence,
one is free to modify the Hilbert-space formulation for a given indivisible
stochastic process as needed, much like changing from one gauge choice
to another in a gauge theory, or like adding physically meaningless
variables to the Lagrangian formulation of a Newtonian system.

With this motivation in place, recall again the basic stochastic\textendash quantum
dictionary \eqref{eq:DefDictionary}: 
\begin{equation}
\stochasticmatrix_{ij}{\left(t\from0\right)}=\tr\parens{\dynop^{\adj}{\left(t\from0\right)}\projector_{i}\dynop{\left(t\from0\right)}\projector_{j}}.\label{eq:DefDictionaryForDilations}
\end{equation}
 The Hilbert-space formulation expressed by the right-hand side can
be manipulated for convenience, provided that the left-hand side of
the dictionary remains unchanged.

In particular, for any integer $D\geq2$, one can freely enlarge,
or \emph{dilate}, the Hilbert-space formulation to a larger dimension
$N\!D$ by the following dilation transformation, as first explained
in Subsection~\ref{subsec:Kraus-Decompositions-and-Unitary-Time-Evolution}:
\begin{equation}
\eqsbrace{\begin{aligned}\dynop{\left(t\from0\right)} & \mapsto\dynop{\left(t\from0\right)}\tensorprod\idmatrix^{\mathcal{I}},\\
\projector_{i} & \mapsto\projector_{i}\tensorprod\idmatrix^{\mathcal{I}},\\
\projector_{j} & \mapsto\projector_{j}\tensorprod\projector^{\mathcal{I}}_{\gamma}.
\end{aligned}
}\label{eq:DefDilation}
\end{equation}
 Here $\idmatrix^{\mathcal{I}}$ is the $D\times D$ identity matrix
on a new \emph{internal} Hilbert space $\hilbspace_{\mathcal{I}}$,
and $\projector^{\mathcal{I}}_{1},\dots,\projector^{\mathcal{I}}_{D}$
collectively form any PVM on that internal Hilbert space satisfying
the usual conditions of mutual exclusivity, 
\begin{equation}
\projector^{\mathcal{I}}_{\gamma}\projector^{\mathcal{I}}_{\delta}=\delta_{\gamma\delta}\projector^{\mathcal{I}}_{\gamma},\label{eq:InternalProjectorsMutuallyExclusive}
\end{equation}
 and completeness, 
\begin{equation}
\sum^{D}_{\gamma=1}\projector^{\mathcal{I}}_{\gamma}=\idmatrix^{\mathcal{I}}.\label{eq:InternalProjectorsComplete}
\end{equation}
 It is then a mathematical identity that one can rewrite the stochastic\textendash quantum
dictionary \eqref{eq:DefDictionary} as 
\begin{equation}
\eqsbrace{\begin{aligned}\stochasticmatrix_{ij}{\left(t\from0\right)} & =\tr{\left(\tr_{\mathcal{I}}{\left({\left[\dynop^{\adj}{\left(t\from0\right)}\tensorprod\idmatrix^{\mathcal{I}}\right]}{\left[\projector_{i}\tensorprod\idmatrix^{\mathcal{I}}\right]}\right.}\right.}\\
 & \qquad\times{\left.{\left.{\left[\dynop{\left(t\from0\right)}\tensorprod\idmatrix^{\mathcal{I}}\right]}{\left[\projector_{j}\tensorprod\projector^{\mathcal{I}}_{\gamma}\right]}\right)}\right)},
\end{aligned}
\negthickspace\negthickspace\negthickspace\negthickspace}\label{eq:DilatedDictionary}
\end{equation}
 with a second trace, or partial trace, over the internal Hilbert
space $\hilbspace_{\mathcal{I}}$. The choice of value for the label
$\gamma$ here is immaterial, with different choices of $\gamma$
related by gauge transformations.

One can equivalently write the dilated form \eqref{eq:DilatedDictionary}
of the dictionary in \emph{block-matrix form} as 
\begin{equation}
\stochasticmatrix_{ij}{\left(t\from0\right)}=\tr_{\mathcal{I}}{\left({\left[\dynop_{ij}{\left(t\from0\right)}\right]}^{\mathcal{I}\adj}{\left[\dynop_{ij}{\left(t\from0\right)}\right]}^{\mathcal{I}}\projector^{\mathcal{I}}_{\gamma}\right)}.\label{eq:DilationStochasticMatrixDictionaryBlockForm}
\end{equation}
 Here ${\left[\dynop_{ij}{\left(t\from0\right)}\right]}^{\mathcal{I}}$
is a diagonal $D\times D$ matrix consisting of repeated copies of
the specific entry $\dynop_{ij}{\left(t\from0\right)}$ (for fixed
$i,j$) along the diagonal: 
\begin{equation}
{\left[\dynop_{ij}{\left(t\from0\right)}\right]}^{\mathcal{I}}\defeq\dynop_{ij}{\left(t\from0\right)}\,\idmatrix^{\mathcal{I}}.\label{eq:DefDilatedBlockTimeEvOp}
\end{equation}
 Meanwhile, the adjoint operation $\adj$ in \eqref{eq:DilationStochasticMatrixDictionaryBlockForm}
acts on this $D\times D$ block matrix ${\left[\dynop_{ij}{\left(t\from0\right)}\right]}^{\mathcal{I}}$,
so it does not transpose the indices $i$ and $j$ on the $N\times N$
matrix $\dynop_{ij}{\left(t\from0\right)}$ itself: 
\begin{equation}
{\left[\dynop_{ij}{\left(t\from0\right)}\right]}^{\mathcal{I}\adj}\defeq\overconj{{\left[\dynop_{ij}{\left(t\from0\right)}\right]}}^{\mathcal{I}}.\label{eq:DefAdjointDilatedBlockTimeEvOp}
\end{equation}
 It follows that 
\begin{equation}
{\left[\dynop_{ij}{\left(t\from0\right)}\right]}^{\mathcal{I}\adj}{\left[\dynop_{ij}{\left(t\from0\right)}\right]}^{\mathcal{I}}\projector^{\mathcal{I}}_{\gamma}=\verts{\dynop_{ij}{\left(t\from0\right)}}^{2}\projector^{\mathcal{I}}_{\gamma},\label{eq:CalculationBlockMatrixForDilatedDictionary}
\end{equation}
 so the trace over $\hilbspace_{\mathcal{I}}$ indeed yields $\verts{\dynop_{ij}{\left(t\from0\right)}}^{2}=\stochasticmatrix_{ij}{\left(t\from0\right)}$,
as required by \eqref{eq:StochasticMatrixEntryFromAbsValSquare}.

In this dilated version of the Hilbert-space formulation, Schur\textendash Hadamard
gauge transformations \eqref{eq:DefLocalInTimeSchurHadamardGaugeTransformation}
are enhanced to the following local-in-time gauge transformations,
which have not yet been described in the research literature and therefore
constitute another new result: 
\begin{equation}
{\left[\dynop_{ij}{\left(t\from0\right)}\right]}^{\mathcal{I}}\mapsto V^{\mathcal{I}}_{{\left(ij\right)}}{\left(t\right)}{\left[\dynop_{ij}{\left(t\from0\right)}\right]}^{\mathcal{I}}.\label{eq:LocalInTimeBlockwiseUnitaryTransformations}
\end{equation}
 Here $V^{\mathcal{I}}_{{\left(ij\right)}}{\left(t\right)}$ are a
set of $N^{2}$ unitary, $D\times D$ matrices, where each such unitary
matrix as a whole is labeled by a specific pair ${\left(ij\right)}$
of configuration labels: 
\begin{equation}
V^{\mathcal{I}\adj}_{{\left(ij\right)}}{\left(t\right)}=\parens{V^{\mathcal{I}}_{{\left(ij\right)}}{\left(t\right)}}^{-1}.\label{eq:LocalInTimeBlockwiseUnitaryMatrices}
\end{equation}

The gauge transformations \eqref{eq:LocalInTimeBlockwiseUnitaryTransformations}
will not generally preserve the factorization $\dynop{\left(t\from0\right)}\tensorprod\idmatrix^{\mathcal{I}}$
appearing in \eqref{eq:DilatedDictionary}, so they motivate considering
more general $N\!D\times N\!D$ time-evolution operators $\tilde{\dynop}{\left(t\from0\right)}$,
in terms of which the dilated dictionary \eqref{eq:DilatedDictionary}
takes the form 
\begin{equation}
\stochasticmatrix_{ij}{\left(t\from0\right)}=\tr{\left(\tr_{\mathcal{I}}{\left(\tilde{\dynop}^{\adj}{\left(t\from0\right)}{\left[\projector_{i}\tensorprod\idmatrix^{\mathcal{I}}\right]}\tilde{\dynop}{\left(t\from0\right)}{\left[\projector_{j}\tensorprod\projector^{\mathcal{I}}_{\gamma}\right]}\right)}\right)}.\label{eq:DilatedDictionaryGenericTimeEvOp}
\end{equation}
 Any $N\!D\times N\!D$ matrix $\tilde{\dynop}{\left(t\from0\right)}$
appearing on the right-hand side of this dictionary and satisfying
the natural generalization of the summation condition \eqref{eq:SumTimeEvOpAbsValSqEq1}
is guaranteed to lead to a valid transition matrix $\stochasticmatrix_{ij}{\left(t\from0\right)}$
on the left-hand side, so working with a dilated Hilbert-space formulation
essentially provides a larger \textquoteleft canvas\textquoteright{}
for designing transition matrices.

As a simple example of a dilation for the case $D=2$, one can formally
eliminate the complex numbers from a quantum system's Hilbert space
(Myrheim 1999)\nocite{Myrheim:1999qmoarhs}. Specifically, by increasing
the system's Hilbert-space dimension from $N$ to $2N$, one can replace
the imaginary unit $i\defeq\sqrt{-1}$ with the real-valued $2\times2$
matrix ${\left(\begin{smallmatrix}0 & -1\\
1 & 0
\end{smallmatrix}\right)}$, with the enhanced version \eqref{eq:LocalInTimeBlockwiseUnitaryTransformations}
of Schur\textendash Hadamard gauge transformations now consisting
of two-dimensional rotations of the internal Hilbert space $\hilbspace_{\mathcal{I}}$.\footnote{Importantly, one can prove the uncertainty principle just as well
with the imaginary unit $i$ represented by a $2\times2$ matrix in
this way. } One can then represent the complex-conjugation operator $K$ appearing
in \eqref{eq:AntiUnitaryDynamicalSymmetryUsingComplexConjOp} as the
real-valued $2\times2$ matrix ${\left(\begin{smallmatrix}0 & 1\\
1 & 0
\end{smallmatrix}\right)}$. The result is that all unitary and anti-unitary operators become
$2N\times2N$ real orthogonal matrices. One cost of using this \textquoteleft real\textquoteright{}
representation, however, is that the Hilbert spaces of composite systems
will not factorize as neatly into Hilbert spaces for their constituent
subsystems.

As a much more significant application of dilations, recall that any
transition matrix $\stochasticmatrix_{ij}{\left(t\from0\right)}$
has a Kraus decomposition \eqref{eq:StochasticMatrixFromKrausDecomposition},
which one can equivalently write as 
\begin{equation}
\stochasticmatrix_{ij}{\left(t\from0\right)}=\sum^{N}_{\beta=1}\tr\parens{\krausmatrix^{\adj}_{\beta}{\left(t\from0\right)}\projector_{i}\krausmatrix_{\beta}{\left(t\from0\right)}\projector_{j}}.\label{eq:DictionaryFromKrausDecompositionForDilation}
\end{equation}
 As explained in Subsection~\ref{subsec:Kraus-Decompositions-and-Unitary-Time-Evolution}
and proved in other work (Barandes 2025, 2023)\nocite{Barandes:2025tsqc,Barandes:2023tsqt},
the Stinespring dilation theorem~(Stinespring 1955, Keyl 2002)\nocite{Stinespring:1955pfoc,Keyl:2002foqit}
then guarantees that by an appropriate dilation to a larger Hilbert
space if necessary, one can express $\stochasticmatrix_{ij}{\left(t\from0\right)}$
in terms of a unitary time-evolution operator $\tilde{\timeevop}{\left(t\from0\right)}$:
\begin{equation}
\stochasticmatrix_{ij}{\left(t\from0\right)}=\tr{\left(\tr_{\mathcal{I}}{\left(\tilde{\timeevop}^{\adj}{\left(t\from0\right)}{\left[\projector_{i}\tensorprod\idmatrix^{\mathcal{I}}\right]}\tilde{\timeevop}{\left(t\from0\right)}{\left[\projector_{j}\tensorprod\projector^{\mathcal{I}}_{\gamma}\right]}\right)}\right)}.\label{eq:DictionaryAfterDilation}
\end{equation}

As yet another key application of dilations, a dilated Hilbert-space
formulation can make it possible to describe new kinds of emergeables.
Some of these \emph{dilation-emergeables} may be observables that
can yield definite results in measurement processes, along the lines
described in other work (Barandes 2025)\nocite{Barandes:2025tsqc},
despite not having a direct meaning solely at the level of the system's
underlying configuration space.

In this way, an indivisible stochastic process based on a configuration
space can easily accommodate emergent observables that describe empirically
meaningful patterns in the dynamics and that model all kinds of quantum
phenomena. Indeed, obtaining a unitary time-evolution operator for
a given system may \emph{require} dilating the Hilbert space in just
this way, as in \eqref{eq:DictionaryAfterDilation}.

It is important to keep in mind that whether or not one actually carries
out this formal dilation of the Hilbert-space formulation, the stochastic
dynamics of the underlying indivisible stochastic process will still
be the same. Any emergent patterns in the system's stochastic dynamics
that are made manifest or explicit by the dilation, as represented
by any new dilation-emergeables that arise, were always there all
along, albeit in a non-manifest or implicit way.

An important potential example of this last application is intrinsic
spin. If one wished to introduce spin as a dilation-emergeable, then
one could merely dilate the Hilbert space to $N\!D$ dimensions, introduce
a $D$-dimensional representation of $SO{\left(3\right)}$ for the
internal Hilbert space, and then require that the dilated time-evolution
operator had the appropriate form of rotation symmetry. This approach
to representing spin would ensure that despite picking an arbitrary
three-dimensional coordinate axis in the process of formally carrying
out the dilation of the Hilbert space\textemdash such as by choosing
the spin-$z$ operator to be diagonal on the dilated Hilbert space\textemdash the
underlying indivisible stochastic process would not fundamentally
involve any preferred direction or entail any basic violation of rotation
invariance.

\section{Discussion and Future Work\label{sec:Discussion-and-Future-Work}}

The present paper discussed the basic theory of stochastic processes,
with an emphasis on generalizations to accommodate non-Markovianity,
before introducing the notion of an indivisible stochastic process.
The paper then reviewed the stochastic\textendash quantum correspondence
between indivisible stochastic processes and quantum systems, leading
to the indivisible interpretation of quantum theory, or indivisible
quantum theory (Barandes 2025)\nocite{Barandes:2025tsqc}. This interpretation
has a thoroughly realist orientation, and does not entail parallel
universes, nor does it involve perspectival or relational notions
of ontology.

The axioms of indivisible quantum theory are simpler to state and
more physically transparent than the Dirac\textendash von Neumann
axioms of textbook quantum theory.
\begin{itemize}
\item Kinematical axiom: For each model under consideration, one picks an
appropriate configuration space, whose members are the possible configurations
of the system being modeled, treated as elementary according to the
model. The configuration space is a fixed feature of the model, meaning
that it does not vary between real-world runs or instantiations of
the model.
\item Dynamical axiom: For arbitrary target times, and for conditioning
times corresponding to division events, the model's dynamical laws
consist of transition probabilities that take the form of conditional
probabilities for the system to be in a particular configuration at
each target time, given that the system is in a particular configuration
at each conditioning time. Division events may occur naturally within
the model's own dynamics, and can also be generated spontaneously
through interactions with other systems. For example, division events
are generated during a measurement process, which can be modeled as
just another stochastic process. At the level of the given model,
the dynamical laws are fixed features.
\item Epistemic axiom: The system has some time-dependent standalone probability
distribution to be in a particular configuration at any given target
time. This standalone probability distribution is connected between
different times by the model's transition probabilities, and is contingent,
meaning that it can vary between runs of the model.
\end{itemize}
It is worth noting that the kinematical and epistemic axioms here
are essentially classical, in the sense that they involve classical
notions of ontology and probability. The distinctly non-classical
ingredient is the dynamical axiom, which replaces the effectively
Markovian differential equations of classical theories with dynamical
laws that consist of a sparse set of indivisible transition probabilities.

Notice that the configurations of an indivisible stochastic process
play the role of what have historically been called \textquoteleft hidden
variables.\textquoteright{} However, they are not hidden in a literal
sense, because they are what one actually sees in experiments, and
they do not supplement or augment the traditional wave function, but
instead replace wave functions as the ontological ingredients of quantum
theory.

Any mention of hidden variables may immediately bring to mind a number
of no-go theorems about non-locality, including the various versions
of Bell's theorem (Bell 1964, 1976, 1990)\nocite{Bell:1964oeprp,Bell:1976ttolb,Bell:1990lnc}.
These theorems will be addressed in detail in future work. In addition
to exploring implications for how to think about causation at a microphysical
level, now that the basic dynamical laws are no longer Markovian differential
equations, that other work will also argue that locality in space
is preserved at the cost of non-Markovianity. One can view non-Markovianity
roughly as a form of non-locality in time that is consistent with
the light-cone structure of special relativity, and which is arguably
unavoidable in quantum theory anyway, as pointed out by others (Glick,
Adami 2020)\nocite{GlickAdami:2020manmqm}.

Future work will extend the analysis of symmetries and dilations to
more examples. At a broader level, it would be interesting to examine
what else this new formulation of quantum theory can teach us about
laws, probability, causation, as well as explore new applications
for quantum computing, and new ways of thinking about generalizing
quantum theory to accommodate gravity.

\section*{Acknowledgments}

The author would especially like to acknowledge Emily Adlam, David
Albert, Howard Georgi, David Kagan, and Logan McCarty for extensive
discussions during the writing of this paper. The author would also
like to thank Scott Aaronson, David Baker, Francesco Buscemi, Ignacio
Cirac, Ned Hall, David Kaiser, Serhii Kryhin, Barry Loewer, Alex Meehan,
Xiao-Li Meng, Simon Milz, Kavan Modi, Wayne Myrvold, Filip Niewinski,
Jill North, Jamie Robins, Noel Swanson, Vivishek Sudhir, Xi Yin, and
Nicole Yunger Halpern for helpful conversations.

\section*{Declarations}

\paragraph{Funding:}

This work was supported by Harvard University, the American Institute
of Mathematics, the Eutopia Foundation, and the Institute of Art and
Ideas.

\paragraph{Conflicts of Interest and Data Availability:}

The author has no conflicts of interest to report or any data to make
available.

\bibliographystyle{1_home_jacob_Documents_Work_My_Papers_2023-Stoc___ses_and_Quantum_Theory_custom-abbrvalphaurl}
\bibliography{0_home_jacob_Documents_Work_My_Papers_Bibliography_Global-Bibliography}

\end{document}